\documentstyle[aps,prl,epsf]{revtex}

\newcommand{\be}{\begin{equation}}
\newcommand{\ee}{\end{equation}}
\newcommand{\bea}{\begin{eqnarray}}
\newcommand{\eea}{\end{eqnarray}}
\newcommand{\bc}{\begin{center}}
\newcommand{\ec}{\end{center}}
\def\lsim{\raise0.3ex\hbox{$<$\kern-0.75em\raise-1.1ex\hbox{$\sim$}}}
\def\gsim{\raise0.3ex\hbox{$>$\kern-0.75em\raise-1.1ex\hbox{$\sim$}}}

\begin{document}
\pagestyle{plain}
\title{Sea Quark Effects on Quarkonia}

\author{
CP-PACS Collaboration  \\
$^{1,5}$T.~Manke, $^1$A.~Ali Khan, $^2$S.~Aoki, $^{1,2}$R.~Burkhalter, $^1$S.~Ejiri, $^3$M.~Fukugita,
$^4$S.~Hashimoto, $^{1,2}$N.~Ishizuka, $^{1,2}$Y.~Iwasaki, $^{1,2}$K.~Kanaya, $^1$T.~Kaneko,  
$^4$Y.~Kuramashi, $^1$K.~Nagai,
$^4$M.~Okawa,  $^1$H.~P.~Shanahan, $^{1,2}$A.~Ukawa, $^{1,2}$T.~Yoshi\'e}

\address{
$^1$Center for Computational Physics,
University of Tsukuba, Tsukuba, Ibaraki 305-8577, Japan \\
$^2$Institute of Physics, University of
Tsukuba, Tsukuba, Ibaraki 305-8571, Japan \\
$^3$Institute for Cosmic Ray Research,
University of Tokyo, Tanashi, Tokyo 188-8502, Japan \\
$^4$High Energy Accelerator Research Organization
(KEK), Tsukuba, Ibaraki 305-0801, Japan \\
$^5$Department of Physics, Columbia University, New York, NY 10027, USA}

\date{\today}

\maketitle
\vspace*{-5.5cm}
\noindent
%\hfill \mbox{UTCCP-P-54}
\vspace*{5.5cm}

\begin{abstract}
We study the effects of two dynamical sea quarks on the spectrum of heavy
quarkonia. Within the non-relativistic approach to Lattice QCD we found
sizeable changes to the hyperfine splitting, but we could not observe any changes
for the fine structure. We also investigated the scaling behaviour of our
results for several different lattice spacings.
\end{abstract}

\section{Introduction}

The experimental efforts to pin down the parameters of the Standard Model
have been paralleled by intense theoretical attempts to provide
experimentalists with non-perturbative constraints from Quantum Chromodynamics (QCD).
It is hoped that Lattice QCD will ultimately provide such important information.
To this end it is crucial to understand and control the systematic errors in
numerical calculations, which rely on extrapolations and interpolations to the physical point.
This important task is particularly demanding for heavy quark phenomenology,
where one has to describe accurately both the light and heavy quarks in the system.

In particular the inclusion of light dynamical fermions in the gluon background is still a daunting task and requires the largest fraction of computational resources.
In the past these restrictions led to the so-called {\it quenched}
approximation, in which only {\it valence quarks} are allowed to propagate 
in a purely gluonic background, whereas the virtual creation of {\it sea quarks} is ignored.
We have shown in a recent work \cite{cppacs_quenched} that this results in systematic deviations 
in the lattice prediction of the light hadron spectrum 
from the observed experimental data.
More recently it has also been found that the inclusion of two dynamical sea quarks 
has a significant effect on the light hadron spectrum and quark masses \cite{cppacs_nf2,kaneko}.
This is of course analogous to QED, 
% new 
where the inclusion of all higher order effects,
% old
% where inclusion of vacuum polarisation effects, 
which could be made through perturbation theory, 
resulted in an impressive agreement with
experimental observations. A distinctive aspect of QCD is that 
a proper non-perturbative treatment is in
order so as to provide high-precision tests of this theory.
Here we take this as our motivation to study heavy quarkonia in the presence (and
absence) of dynamical sea quarks.
 
Heavy quark systems have long been considered an ideal testing ground for QCD
and they have triggered the development of static potential models \cite{WEF} and heavy
quark effective field theories \cite{HQET}. On the lattice, heavy quarks have frequently
been studied using a non-relativistic approach to QCD (NRQCD
\cite{heavy_bound}) or a relativistic formulation promoted by the Fermilab
group \cite{fermilab}. Both provide effective descriptions to deal with large
scale differences, which are difficult to accommodate on conventional lattices.
NRQCD has been quite successful in reproducing the spin-independent spectrum of
heavy quarkonia \cite{nrqcd_first} owing to the fact that the quarks within such mesons move with small velocities $v$ such that $v^2 \ll c^2$.  

As an effective field theory the predictive power of NRQCD relies on
the control of higher dimensional operators, which have to be matched to the
non-relativistic expansion of QCD. This has been the subject of many previous
studies \cite{nrqcd_improved,morningstar,omv6}. As a result of these works it seemed
plausible that bottomonium states could be accurately described in the NRQCD
approach, whereas the spin structure of charmonium is very sensitive to the
higher order relativistic corrections \cite{lidsey,trottier}. 

Within the NRQCD framework sea quark effects on the heavy quarkonium spectra have previously been 
studied at a lattice spacing of $a\approx 0.1$ fm 
using the Kogut-Susskind \cite{nrqcd_alphas} or the Wilson \cite{spitz}
action for sea quarks. 
Here we present results for three lattice spacings in the range
$a\approx 0.2-0.1$ fm, paying particular attention to the 
dependence on the sea quark mass and scaling properties.  
Our gauge configurations are generated with 
a renormalization-group (RG)-improved gluon action \cite{iwasaki} 
and a tadpole-improved 
clover quark action \cite{SW} for two dynamical flavours \cite{cppacs_nf2}.  
Some measurements are also made on quenched configurations 
generated with the same gluon action for making direct comparisons 
of dynamical and quenched results.

In Section \ref{sec:actions} we introduce the actions which we use in our calculation.
In Section \ref{sec:simulation} we give the details of our simulation, meson operators and fitting
methods. Our results are discussed in Section \ref{sec:results} and Section
\ref{sec:conclusion} concludes this paper.

\section{Actions}
\label{sec:actions}
\subsection{Glue: RG-Improved Action}
Since there is no unique discretisation of the continuum gluon action
one can employ a set of operators to cancel some of the discretisation errors
in the lattice action. The most common choice is to simply add a
rectangular $1\times 2$ plaquette, ${\rm Tr} R_{\mu \nu}$, to the standard
Wilson action, ${\rm Tr} P_{\mu \nu}$,
\be
S_{\rm g}(g^2) = \frac{1}{g^2}\left\{c_0 {\rm Tr} P_{\mu \nu} + c_1 {\rm Tr} R_{\mu \nu}\right\}~~,
\ee
where ${\rm Tr}$ denotes the trace over all indices and $c_0 + 8 c_1 = 1$. All such choices have the same continuum limit, but different discretisation 
errors. 
Here we adopt a prescription which is motivated by an RG-analysis
of the pure gauge theory ($c_1=-0.331$ \cite{iwasaki}).
This has proven to be a suitable choice compared to, say, 
the Symanzik-improved action ($c_1=-1/12$), for reducing scaling violation on coarse lattices. 
Instead of the coupling constant squared, $g^2$, we often quote 
the value of $\beta \equiv 6/g^2$.

\subsection{Light Quarks: Clover Action}
The standard discretisation of the fermion action removes the doublers at the
expense of ${\cal O}(a)$ discretisation errors. It is possible to remove these
errors by adding a single operator (the clover term) as first suggested in
\cite{SW}:

\be
S_{\rm q}(g^2,m_q) = \bar q~Q~q= \bar q~(\slash \hskip -7pt \Delta + m_q )~q + a r~\bar q~
\Delta^2~q -c_{sw}(g^2)~ar~\frac{ig}{4} \bar q~ \sigma_{\mu\nu}F_{\mu \nu}~q~~.
\label{eq:light_fermion}
\ee
Here the second term removes the doublers a la Wilson and the last is to reduce the
resulting ${\cal O}(a)$ errors. 
We choose $r=1$ and $c_{sw}= (1 - 0.1402~g^2)^{-3/4}$. 
For the latter we follow the tadpole prescription of \cite{viability_pt},
which has been shown to improve the convergence  of lattice perturbation theory
significantly. Our choice is based upon the perturbative plaquette values as 
determined in \cite{iwasaki}.  To one-loop order our choice differs from the correct 
value \cite{aoki} only by $0.008g^2$. 
Hence we expect only small ${\cal O}(\alpha a)$  scaling violations 
due to radiative corrections from the clover action,  and 
${\cal O}(a^2)$ errors from the gluon action.

%After integration over the Grassmann variables $q$ and $\bar
%q$ we are left with the determinant of the fermion matrix, 
%which is evaluated stochastically.
In our simulations we work with two flavours of degenerate quarks of 
a common mass: $m_q=m_u=m_d$.
For further reference, it is customary to replace the bare quark mass by the hopping parameter: $\kappa \equiv 1/2(m_qa + 4)$.
Since the direct simulation of realistic light Wilson quarks is not feasible on
present-day computers we study the spectrum at a sequence of different $\kappa$.

\subsection{Heavy Quarks: NRQCD}
With the above actions we generated full QCD dynamical configurations on lattices of about 2.5 fm in spatial extent and lattice
spacings ranging from approximately 0.1 to 0.2 fm. Such lattices are particularly suited
to study light quark physics which is determined by a single scale:
$\Lambda_{QCD} \approx 200$ MeV. However, for systems containing slow-moving and
heavy quarks we have to adjust the theoretical description to take into account all the
non-relativistic scales: mass ($m_Q$), momentum ($m_Qv$) and kinetic energy ($m_Qv^2$).
In this work we implement the NRQCD formulation to propagate the heavy quarks in a
given gluon background. This approach has met with considerable success for
b-quarks \cite{nrqcd_improved,omv6}. Also charm quarks have previously been studied
in this framework \cite{lidsey,trottier}.

Whereas the relativistic boundary value problem requires several iterations to 
determine the quark propagator, the NRQCD approach has the numerical advantage to solve the 
two-spinor theory as a much simpler initial value problem.
The forward propagation of the source vector, $S({\bf x})$, is described by:

\bea
G({\bf x},t+1) &=& \left(1-\frac{a}{2}\delta H \right)\left(1-\frac{aH_0}{2n}\right)^n U_t^{\dag}(x) \left(1-\frac{aH_0}{2n}\right)^n
\left(1-\frac{a}{2}\delta H \right) G({\bf x},t) ~~,~~~t \ge  1~~,\nonumber \\
G({\bf x},1)   &=& \left(1-\frac{a}{2}\delta H \right)\left(1-\frac{aH_0}{2n}\right)^n U_t^{\dag}(x)
\left(1-\frac{aH_0}{2n}\right)^n \left(1-\frac{a}{2}\delta H \right) S({\bf x})~~,~~~t = 0~~,
\label{eq:evol}
\eea
where
\bea
H_0 &=& -  \frac{\Delta^2}{2m_Q} ~~, \nonumber \\
\delta H &=& - c_0 \frac{\Delta^4}{8m_Q^3} - c_1 \frac{1}{2m_Q} \sigma\cdot g{\bf \tilde B}
+ c_2 \frac{i}{8m_Q^2}(\tilde{\Delta} \cdot g{\bf \tilde E} - g{\bf \tilde E} \cdot \tilde{\Delta})
- c_3 \frac{1}{8m_Q^2} \sigma \cdot ( \tilde{\Delta} \times g{\bf \tilde E} -
g{\bf \tilde E} \times \tilde{\Delta})\nonumber \\
& & - c_4 \frac{1}{8m_Q^3}\{\Delta^2,\sigma \cdot g{\bf \tilde B}\}  
    - c_5 \frac{1}{64m_Q^3}\{\Delta^2, \sigma \cdot ( \tilde{\Delta} \times
g{\bf \tilde E} - g{\bf \tilde E} \times \tilde{\Delta})\} 
- c_6 \frac{i}{8m_Q^3} \sigma \cdot g{\bf \tilde E} \times g{\bf \tilde E} \nonumber \\
& & - c_7 \frac{a\Delta^4}{16n m_Q^2} + c_8 \frac{a^2 \Delta^{(4)}}{24m_Q} ~~.
\label{eq:dh}
\eea
The improved lattice operators $\tilde {\Delta}_i, \tilde {\bf E}$ and $\tilde {\bf B}$ are defined as in \cite{nrqcd_improved}.
Other discretisations of NRQCD have been suggested in the past but they
differ only at higher order in the lattice spacing.
Here we follow \cite{omv6,spitz} and employ a formulation which includes all spin terms 
up to $O(mv^6)$ in non-relativistic expansion of QCD. On the coarsest lattice
we checked explicitly that our equation (\ref{eq:evol}) gives the same hyperfine splitting as from the asymmetric version employed in \cite{omv6}.
The parameter $n$ was introduced to stabilise the evolution equation against 
high momentum modes. This is standard in such diffusive problems, but one
should keep in mind that a change in $n$ will have to be accompanied by a
change in $m_Q$ to simulate the same physical system.
Alternatively one could decrease the temporal lattice spacing to prevent the
high momentum modes from blowing up \cite{aniso}.

For a single quark source at point ${\bf y}$ we have $S({\bf x}) = \delta^{(3)}({\bf x,y})$, 
but we also propagate extended objects with the same evolution
equation (\ref{eq:evol}). The operator $H_0$ is the leading kinetic term and $\delta H$ contains
the relativistic corrections.
The last two terms in  $\delta H$ are present to correct for lattice
spacing errors in temporal and spatial derivatives.
For the derivatives we use the improved operators
defined in  \cite{nrqcd_improved} and we also replace the
standard discretized gauge field $F_{\mu \nu}$ by
\be
\tilde{F}_{\mu\nu} = \frac{5}{3}F_{\mu\nu} -
\frac{1}{6}\left(U_{\mu}(x)F_{\mu\nu}(x+\mu) U_{\mu}^{\dag}(x)+
U_{-\mu}(x)F_{\mu\nu}(x-\mu) U_{-\mu}^{\dag}(x) - (\mu
\leftrightarrow \nu)
\right)~~.
\ee

With this prescription we aimed to achieve an accuracy of ${\cal O}(a^4)$ for the heavy 
quark sector. Of course we also expect terms of ${\cal O}(\alpha
a^2)$ due to radiative corrections to this leading order result.
In principle, we have to determine all coefficients in (\ref{eq:dh}) by
(perturbative or non-perturbative) matching to relativistic continuum QCD.
Just as for the light quark sector we rely on a mean-field treatment of all
gauge links to account for the leading radiative corrections.
However, there is no unique prescription for such an improvement and several different 
schemes have been employed in the past. More recently it has been suggested
that the average link variable in the Landau gauge should be less sensitive to
radiative corrections since the gauge fields in Landau gauge have less UV
fluctuations \cite{lepage_u0L}.
Here we adopt this view and divide all links by the appropriate tadpole
coefficient at each value of $(\beta,\kappa)$:
\be
U_{\mu}(x) \to U_{\mu}(x)/u_{0L} ~~,~~ u_{0L} \equiv \frac{1}{3} \langle {\rm Tr} U_{\mu} \rangle_{LG}~~.
\ee
An alternative and gauge-invariant implementation of mean-field improvement
that has frequently been used in the past forces the average plaquette,
$P_{\mu\nu}$, to unity:
\be
U_{\mu}(x) \to U_{\mu}(x)/u_{0P} ~~,~~ u_{0P} \equiv \frac{1}{3} \langle {\rm Tr} P_{\mu\nu} \rangle^{1/4}~~.
\ee
In some selected cases we have also implemented this method to estimate
the effect of unknown radiative corrections to the renormalisation
coefficients, $c_i$. In all applications of (\ref{eq:dh}) we set them to their tree-level value 1.
We denote as ${\cal O}(mv^6,a^2)$ the evolution equation which 
includes all spin-dependent operators up to ${\cal O}(mv^6)$ and where all
operators have been improved to reduce the ${\cal O}(a^2)$ errors.

\section{Simulation}
\label{sec:simulation}
\subsection{Updates, Trajectories and Autocorrelations}
The gauge field configurations with two dynamical sea quarks used for the 
present study  
were generated on the CP-PACS supercomputer at the Center for Computational Physics,
University of Tsukuba. They can be classified by two parameters, $(\beta,\kappa)$, which
determine the lattice spacing and the sea quark mass. 
A standard hybrid Monte Carlo algorithm is used to incorporate the 
effects of the fermion determinant. 
For the matrix inversion we implemented the BiCGStab algorithm.
To reduce the autocorrelations between separate measurements we only used every
fifth or tenth trajectory and binned the remaining data until the statistical
error was independent of the bin size.
In Tab. \ref{tab:simulation} we list the number of trajectories we generated
for each set of couplings along with the other simulation parameters and the
actual number of configurations we used in this study; 
for more details see \cite{cppacs_nf2}.
%Results for the light hadron spectrum have previously appeared in the
%literature which used the same set of configurations \cite{cppacs_nf2}. 
The subsequent determination of the quarkonium  spectrum is subject of this
work. 

Since there has been no previous study of heavy quarkonia using the RG action for the gluon
sector, we also supplemented our calculation by a comparative quenched calculation.  The coupling constant $\beta$ of these quenched configurations were 
chosen so that the string tension of the static quark-antiquark potential 
matches that of one of the dynamical runs.  The parameters of these runs are 
also given in Tab. \ref{tab:simulation}.

\subsection{Meson Operators}
To extract meson masses we calculate two-point functions of operators
with appropriate quantum numbers.
In a non-relativistic setting gauge-invariant meson operators can be
constructed from the two-spinors $\chi^{\dag}(x)$, $\Psi(y)$
and a Wilson line, $W(x,y):~\chi^{\dag}(x) W(x,y) \Psi(y)$. 
The construction of meson states with definite $J^{PC}$ 
from those fundamental operators is standard \cite{mandula} 
(on the lattice $J$ labels the irreducible representations of the octahedral group ($J=A_1,A_2,T_1,T_2,E$)). 

Since here we are only interested in S and P-states it is sufficient to 
consider $\chi^{\dag}(x)\Psi(x)$ and  $\chi^{\dag}(x) \Delta_i\Psi(x)$,
respectively. The corresponding spin-triplets can be constructed by
inserting the Pauli matrices into those bilinears. We also sum over different
polarisations to increase the statistics.

The overlap of those simplistic operators with the state of interest can be
further improved upon. One way is to employ wavefunctions, 
which try to model the ground state more accurately. This requires assumptions 
about the underlying potential and gauge-fixed configurations.
We decided to use a gauge-invariant smearing technique, which regulates the
extent of the lattice operator, with a single parameter $\epsilon$:

\be
\chi^\dag(x)~\hat{O}~\Psi(x) \to \chi^\dag(x)~\hat{O}~(1-\epsilon \Delta^2)^{10}~\Psi(x)~~.
\label{eq:smearing}
\ee
For computational ease we limited this procedure to 10 smearing iterations and 
implemented it only at the source. From such operators we obtain the meson correlators as a Monte Carlo average over all configurations
\be
{\rm C}_\epsilon(x,y) = \langle {\bf tr}\left[G^{\dag}(x,y)~\hat{O}~G(x,y)~(1-\epsilon \Delta^2)^{10}~\hat{O}^\dag~\right] \rangle~~,
\label{eq:correlator}
\ee
where ${\bf tr}$ denotes contraction over all internal degrees of freedom.

For the smeared propagator we solve (\ref{eq:evol}) with
$S({\bf x,y}) = \delta({\bf x,y})~(1-\epsilon \Delta^2)^{10}~\hat{O}^\dag$.
We fix the origin at some (arbitrary) lattice point $y=({\bf y},0)$. This creates a meson
state with all possible momenta. 
In practice we employed up to 8 spatial sources on every configuration.
This is permissible since heavy quarkonia are small compared to the lattice
extent of about 2.5 fm. We explicitly checked the independence of such
measurements by binning.
At the sink,  $x=({\bf x},t)$, we perform a Fourier
transformation to project the correlator onto a given momentum eigenstate:
\be
{\rm C}_{\epsilon,t}({\bf p}) = \sum_{\bf x} C_{\epsilon}({\bf x},t) \exp{(-i {\bf px})}~~.
\label{eq:FT}
\ee
In the trivial case of zero momentum this amounts to simply summing over all
spatial ${\bf x}$.

\subsection{Fitting}
\label{sec:fitting}
\begin{figure}
\begin{center}
\leavevmode
\hbox{\epsfxsize = 12 cm \epsffile{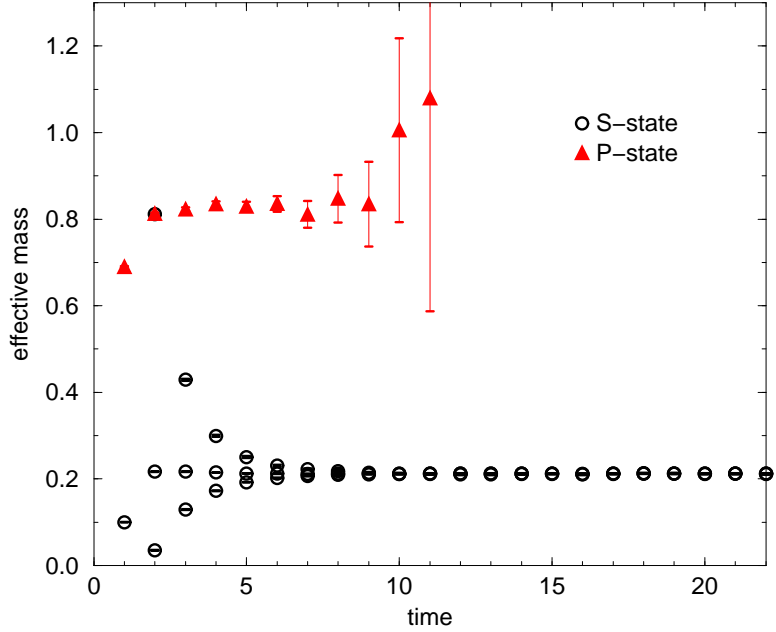}}
\end{center}
\caption{A representative effective mass plot for S- and P-states
at $(\beta,\kappa,m_Q)=(1.80,0.1409,6.00)$. One can clearly observe a plateau for long enough 
times. For the S-states we employed 3 different smearings, which 
result in different overlaps with the ground state. }
\label{fig:effmass}
\end{figure}

Since there is no backward propagation of the heavy quark in our framework, we can fit
the meson propagators to the exponential form: 
\be
y_{\epsilon,t}(a_i,E_i) = \sum_{i=1}^{\rm n_{fit}}~a_i(\alpha , {\bf p}, \epsilon)~
e^{-E_i(\alpha,{\bf p})~t}~~.
\label{eq:theory}
\ee
This is the theoretical prediction for a multi-exponential decay of a state
with momentum ${\bf p}$ and quantum numbers $\alpha$ along Euclidean time $t$.
Different choices for the smearing parameter $\epsilon$ will result in
different overlaps with the ground state and the higher excited states of the 
same quantum number.
In practice it is difficult to project directly onto a given state, so we
chose to extract the ground state from multi-exponential correlated fits.
In some cases we were also able to extract the first excited states 
reliably from our data. 
The simplest way to visualise our data is by means of effective mass plots,
which are expected to display a plateau for long Euclidean times.
In Fig. \ref{fig:effmass} we show a representative plot for one set of simulation
parameters. 

Correlations between different times, $t$, and different smearing radii,
$\epsilon$, of the meson propagator $C_{\epsilon,t}$ are taken into account by employing the
full covariance matrix for the $\chi^2$-minimisation: 
\be
\chi^{2}({\bf \pi}) \equiv
\sum_{r,s=1}(C_r-y_r({\bf \pi}))~\Gamma_{rs}^{-1}~(C_s-y_s({\bf \pi})),~~~~~
\frac{\partial \chi^{2}}{\partial \pi_k}\left({\bf \bar{{\bf \pi}}}\right) = 0~~.
\label{eq:chi2}
\ee
Here, in order to ease the notation,  we introduced multi-indices
($r=[\epsilon,t]$) for the data points and ${\bf \pi}=[a_i,E_i]$ for the parameters.

Our statistical ensembles are large enough to determine the covariance
matrix, $\Gamma_{rs}^{-1}$, with sufficient accuracy. Therefore the inversion of
this matrix did not present a problem. For the spin splittings we applied
(\ref{eq:theory}) also to the ratio of two propagators, $C(\alpha_1)/C(\alpha_2)$. 
In this way we utilized correlations between states of different quantum
numbers to obtain improved estimates for their energy difference.

\begin{figure}
\begin{center}
\leavevmode
\hbox{\epsfxsize = 12 cm \epsffile{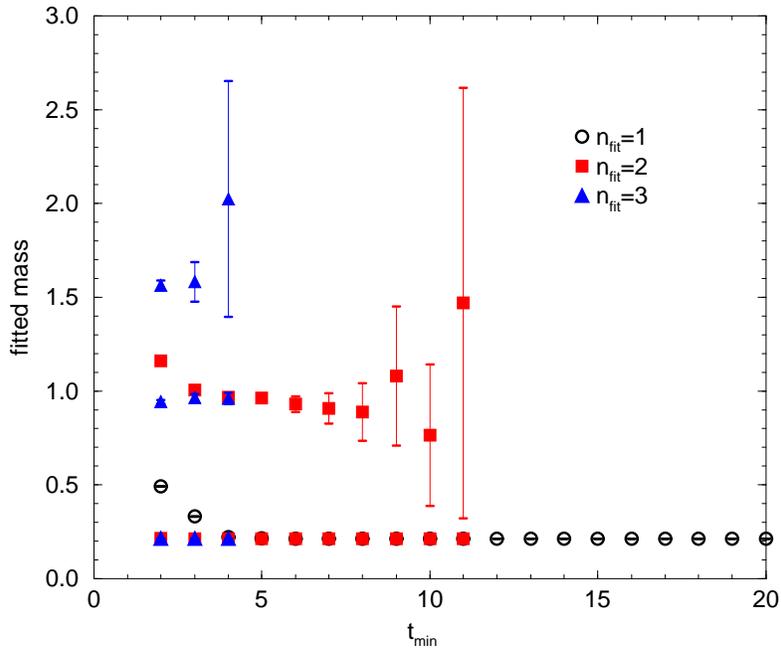}}
\end{center}
\caption{Here we show consistent fit results for the S-state of
Fig. \ref{fig:effmass} when plotted against the start of the fit range, $t_{min}$. 
We fixed $t_{max}=24$  throughout. Different symbols denote different values
for $n_{\rm fit}$ in the multi-exponential fit of Eq. \ref{eq:theory}.}
\label{fig:tmin}
\end{figure}

Statistical fluctuations in the data cause fluctuations in the fit results
determined by (\ref{eq:chi2}). We estimate the covariance matrix,
$\Delta_{kl}$, of the fitted parameters, $\pi_k$, from the inverse of 
$(\partial ^2 \chi^2)/(\partial \pi_k \partial \pi_l)$. We have checked
these errors against bootstrap errors which gave consistent results.
We also require consistent results as we change the fit ranges
$(t_{min},t_{max})$ or the number of exponentials to be fitted.
This is illustrated in Fig. \ref{fig:tmin}, where we show the $t_{min}$-plot
for the S-state of Fig. \ref{fig:effmass}.
The goodness of each fit is quantified by its Q-value \cite{numrec} and we
demand an acceptable fit to have $Q> 0.1$.
Finally we subjected those results to a binning analysis, which takes into account 
autocorrelations of the same measurement at different times in the update process.

\section{Results}
\label{sec:results}
We now present our new results for elements of the spectrum of
heavy quarkonia.
Our data from two quenched lattices is given in Tab. \ref{tab:quenched}
and the dynamical data is collected in Tabs. \ref{tab:ups_beta180} to \ref{tab:charm_beta180}. For notational ease we use dimensionless lattice quantities 
throughout the remainder of this paper, unless stated otherwise.
To convert the lattice predictions into dimensionful quantities we take
the experimental $1P-1S$ splitting to set the scale.

\subsection{Heavy Mass Dependence and Kinetic Mass}
For a given value of the gauge coupling (lattice spacing)
and the mass of the two degenerate sea quarks there is only
one remaining parameter to choose -- the mass of the heavy valence quark.
On the lattice we are free to simulate every arbitrary value,
but in order to obtain physical results we tune the ratio $M_{\rm kin}/(1P-1S)$ 
of the kinetic mass of a quarkonium state and the $1P-1S$ mass splitting 
to its experimental value. The determination of $1S$ and $1P$ masses 
has already been described in Sec. \ref{sec:fitting}. 
The kinetic mass, $M_{\rm kin}$, is defined through the dispersion relation of the quarkonium state:
\be
E({\bf p}) - E({\bf 0}) = \frac{{\bf p}^2}{2M_{\rm kin}} + \ldots ~~~,~~~
{\bf p} = \frac{2 \pi}{L}(n_1,n_2,n_3)~~.
\label{eq:dispers}
\ee

\begin{figure}
\begin{center}
\leavevmode
\begin{tabular}{cc}
\hbox{\epsfxsize = 9 cm \epsffile{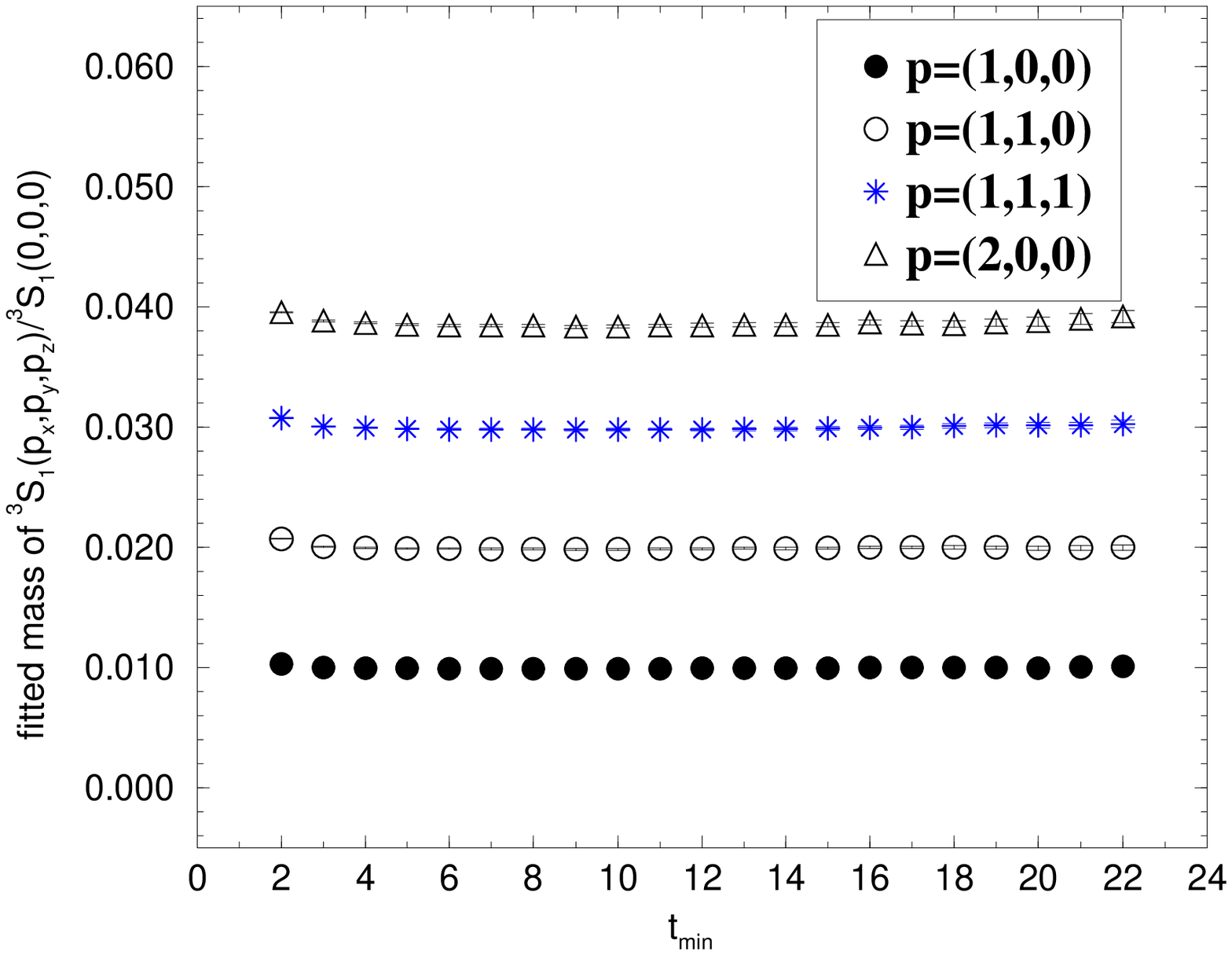}} &
\hbox{\epsfxsize = 9 cm \epsffile{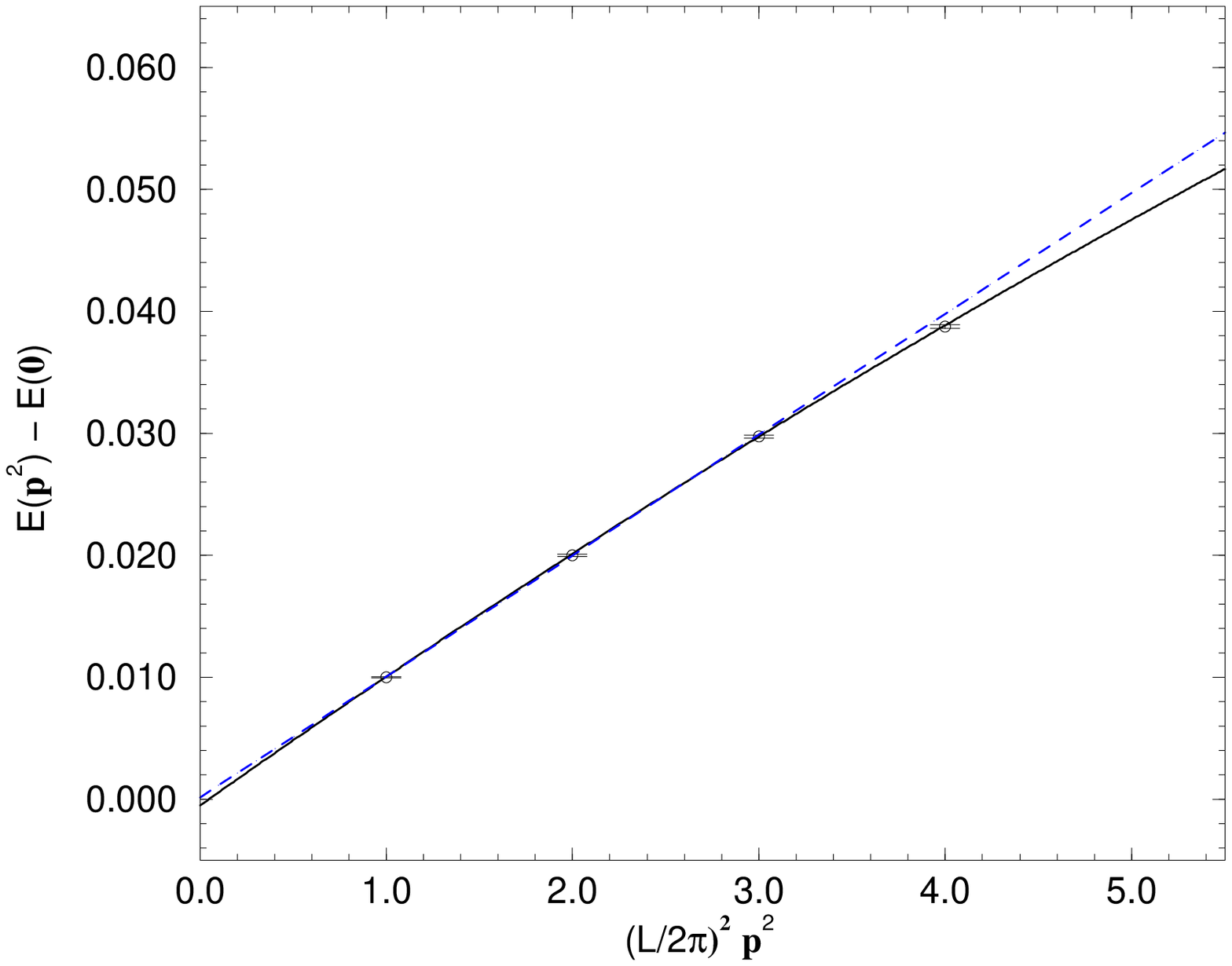}}
\end{tabular}
\end{center}
\caption{These figures illustrate the tuning of the quark mass as described
in the main text. On the left-hand side we show the $t_{min}$-plots for the ratio
fits of different momentum states with respect to the ${^3}S_1$ at rest.
We can perform two consistent fits up to ${\bf p}^2$ (dashed line) and up to
${\bf p}^4$ (solid line) in the dispersion relation, Eq. \ref{eq:dispers}.} 
\label{fig:dispers}
\end{figure}

\begin{figure}
\begin{center}
\leavevmode
\hbox{\epsfxsize = 12 cm \epsffile{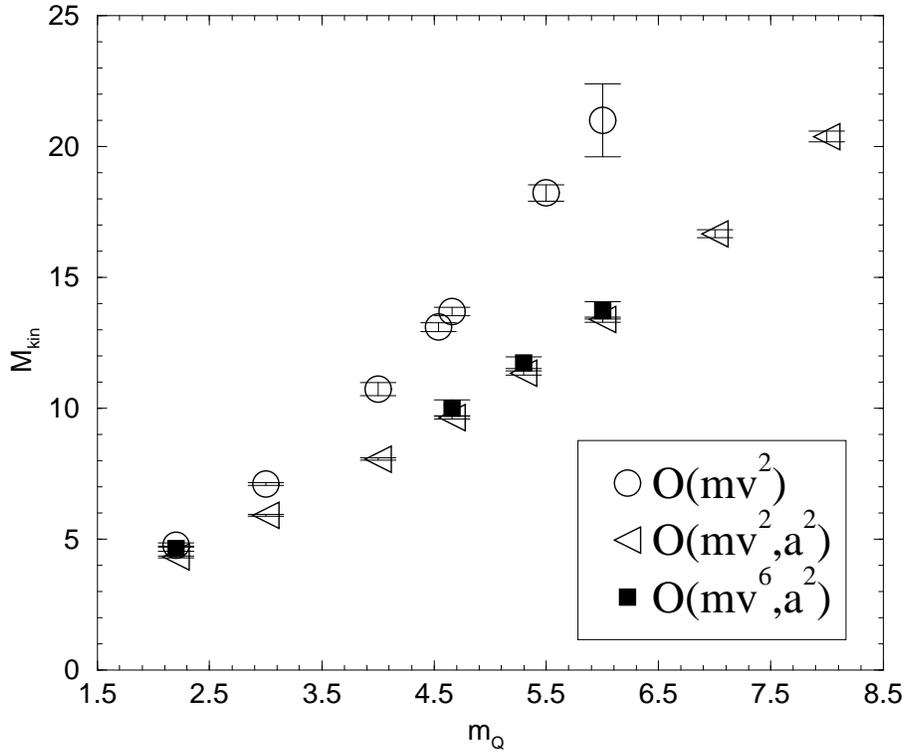}}
\end{center}
\caption{Quark mass dependence of $M_{\rm kin}$.
Here we can see sizeable discretisation errors for almost the whole range
of quark masses between charm and bottom at $(\beta,\kappa)=(1.80,0.1409)$.
The implementation of $O(a^2)$ improvement in the NRQCD approach is
clearly important on our lattices. In contrast, the sensitivity
of $M_{\rm kin}$ to the relativistic correction terms is much smaller.}
\label{fig:aMkin}
\end{figure}

For each heavy quark mass, $m_Q$, we project the  ${^1}S_0$-state 
and the ${^3}S_1$-state onto 5 different momentum 
eigenstates: $(n_1,n_2,n_3) = (0,0,0); (1,0,0);(1,1,0);(1,1,1)$ 
and $(2,0,0)$. We obtain $E({\bf p})-E({\bf 0})$ from ratio fits
and determine the kinetic mass by fitting the data to the dispersion relation.
To this end we have also included higher terms in the expansion of
(\ref{eq:dispers}) and find consistent results for $M_{\rm kin}$.
However, such fits tend to have larger errors and the coefficient of 
${\bf p}^4$ is not well determined.
For better accuracy we normally restrict ourselves to a linear 
fit in ${\bf p}^2$ for the lowest four momenta.
An example of this procedure is given in Fig. \ref{fig:dispers}.  
We have plotted the fitted values of $M_{\rm kin}$ against the heavy quark
mass in Fig.\ref{fig:aMkin}. Large discretisation errors can be seen for
almost all masses, but once we include all $O(a^2)$ correction terms in (\ref{eq:evol}),
the discrepancy due to higher order relativistic corrections is small.
Comparing the relative changes in Fig. \ref{fig:dispers} due to $O(a^2)$ improvement 
at different momentum scales, we can also estimate the size of higher 
order corrections and we expect them to be small.

In this paper we tuned the bare quark mass on all our lattices
$(\beta,\kappa)$, so as to reproduce the experimental value of the 
mass of $\Upsilon$ ($M_{\rm kin}=9.46$ GeV). 
In some selected cases we interpolated the spectrum to this physical point.

\subsection{Scale Determination and $1P-1S$ Splitting}

\begin{figure}
\begin{center}
\leavevmode
\hbox{\epsfxsize = 12 cm \epsffile{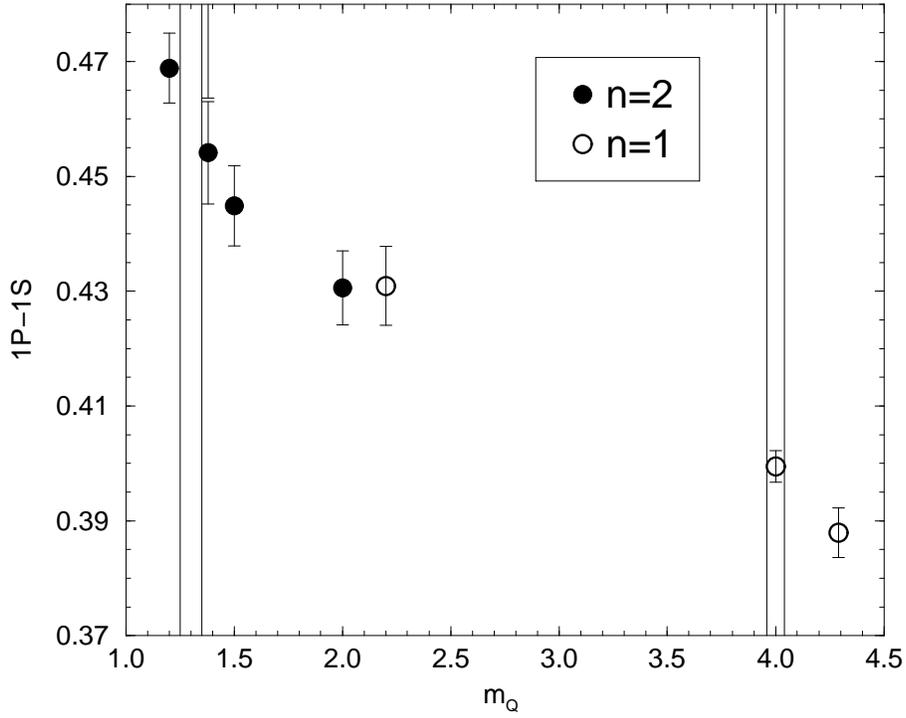}}
\end{center}
\caption{Heavy quark mass dependence of the $1P-1S$ splitting.
We plot the (1P-1S) splitting against the heavy quark mass at
$(\beta,\kappa)=(1.95,0.1375)$ and with two value of the stability
parameter, $n=1,2$. The vertical lines denote the regions of the charmonium and
bottomonium system.}
\label{fig:1P-1S_vs_am}
\end{figure}

It has been noticed in the past that the tuning of $m_Q$ can be done
efficiently since the spin-averaged splitting is not very sensitive to the
quark mass parameter. However, with our newly achieved accuracy we could also
resolve a slight mass dependence of $1P-1S$ in the range from charmonium to the
bottomonium system as shown in Fig. \ref{fig:1P-1S_vs_am}.
% New
The experimental values for the $1P-1S$ splitting show a 4\% decrease
when going from charmonium (458 MeV) to bottomonium (440 MeV), which should 
be compared to a reduction of about 10\% from our simulation
at $N_f=2$ and an unphysical sea quark mass. This larger change is related 
to the running of the strong coupling between the two scales, which
still does not fully match the running coupling in nature.
It is expected that the modified short-range potential will result in a ratio
$(1P-1S)_{c\bar c}/(1P-1S)_{b\bar b}$ bigger than in experiment.

% Old
% This may not be unexpected since also the experimental value of $1P-1S$ in
% charmonium (458 MeV) is larger than the one in bottomonium (449 MeV).  
% Moreover, since we show data with unphysical sea-quark mass we should 
% also expect the ratio of $(1P-1S)_{c\bar c}/(1P-1S)_{b\bar b}$ to be even 
% larger than in experiment, because the strong coupling between 
% those two scales does not run as in nature.  
% Indeed we observe a 10\% increase in Fig. \ref{fig:1P-1S_vs_am} 
% as compared to that of 2\% from experiment.  

\begin{figure}
\begin{center}
\leavevmode
\hbox{\epsfxsize = 12 cm \epsffile{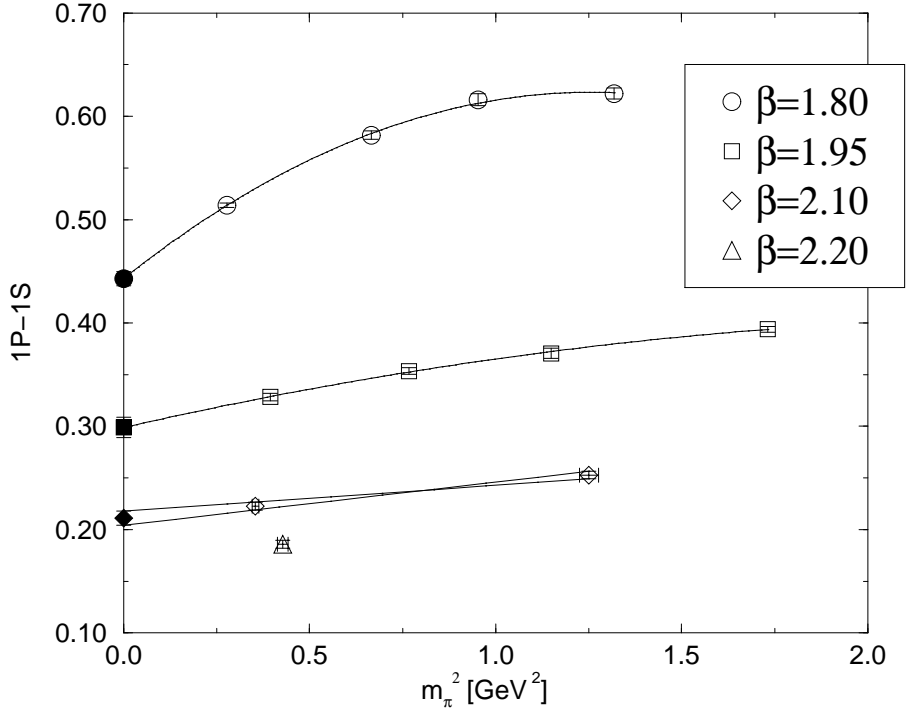}} 
\end{center}
\caption{Light quark mass dependence of $1P-1S$ splitting in Bottomonium. 
We use quadratic fits in $m_\pi^2$ to extrapolate our data from four
different sea quark masses to the chiral limit.
For the two sea quark masses at $\beta=2.10$ we show an estimate of the
chiral limit by drawing straight lines. The single point at $(\beta,\kappa)=(2.20, 0.1368)$ is
shown for completeness.} 
\label{fig:1P-1S_vs_ampi2}
\end{figure}

While the heavy quark mass can be tuned to its physical value as
described in the previous section, this is not possible for
the light quark mass and one has to rely on extrapolations to 
realistic quark masses, where the ratio $m_\pi/m_\rho$ equals the experimental value.
Here we are mainly interested in the behaviour of physical quantities as we 
approach the chiral limit. We take $m_\pi^2$ 
as a measure of the light quark mass and extrapolate quadratically in this parameter.
% NEW
This is a common procedure but we will demonstrate below
that the physical dependence on the sea quark mass may indeed be
difficult to disentangle from unphysical scaling violations.
In taking the naive chiral limit we hope to account for at least
a fraction of the spectral changes towards smaller sea quark masses.
At $\beta=2.10$ we only have results from two values of $\kappa$ and take a
linear estimate for the chiral limit. The chiral behaviour of the $1P-1S$ splitting
is shown in Fig. \ref{fig:1P-1S_vs_ampi2} for all values of $\beta$ in our study.

\begin{figure}
\begin{center}
\leavevmode
\begin{tabular}{c}
\hbox{\epsfxsize = 15 cm \epsffile{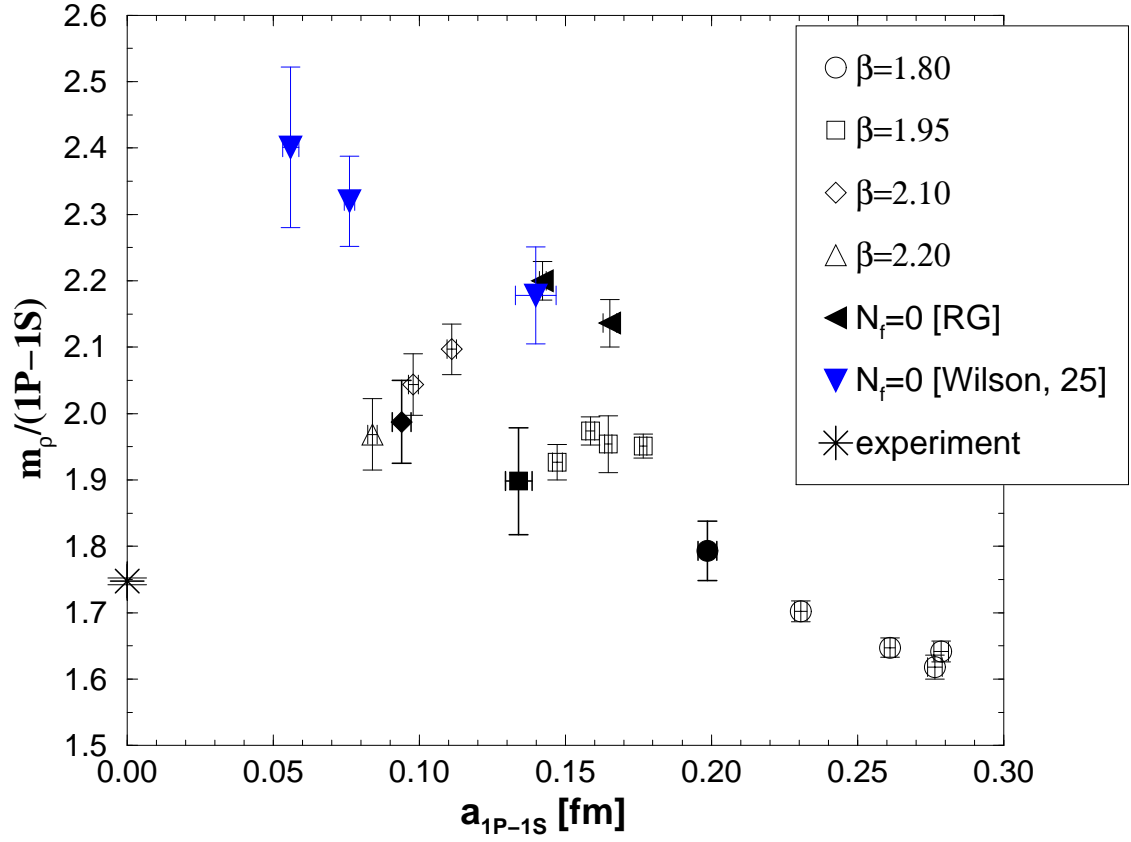}}
\end{tabular}
\end{center}
\caption{Here we show the ratio $m_\rho/(1P-1S)$, where scaling violations can be seen.
In each case we use open symbols to denote data from dynamical calculations
with different sea quark mass and full symbols to mark the chirally extrapolated
values. Representative quenched results are also shown as full symbols. }
\label{fig:mrho_1P-1S}
\end{figure}

In quenched simulations there exist uncertainties when setting the scale 
from different physical quantities.  We expect these uncertainties to be 
reduced in our simulations incorporating two light dynamical flavours. 
To examine this point we compare in
Fig. \ref{fig:mrho_1P-1S} our results for $1P-1S$ splitting with the data 
for $m_\rho$ as a
representative example from the light quark sector. If it were not for 
quenching effects and lattice spacing artefacts, one would expect the ratio
$m_\rho/(1P-1S)$ to equal its experimental value.

It is encouraging to see that the dynamical calculations are
always and significantly closer to the experimental value of 1.75 than 
the corresponding quenched simulations. 
This demonstrates the importance of dynamical over quenched simulations.
The scaling violations in this ratio do not fully cancel, however; 
we observe a 10\% shift in $m_\rho/(1P-1S)$ over $a\approx 0.2-0.1$ fm.   
Keeping in mind that we are working on rather coarse lattices with 
$a\gsim 0.1$ fm, 
the remaining scaling violations are perhaps not too surprising. 

\begin{figure}
\begin{center}
\leavevmode
\hbox{\epsfxsize = 12 cm \epsffile{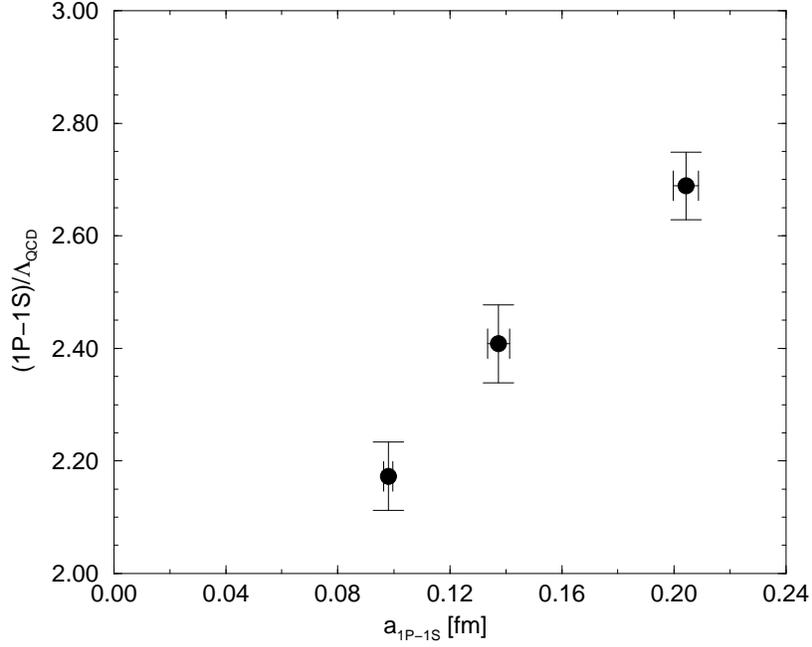}}
\end{center}
\caption{Asymptotic scaling. In this plot we take the chirally extrapolated
values for $1P-1S$ and compare their scaling behaviour with respect to
$\Lambda_{\rm QCD}$. The latter is taken from 2-loop perturbation theory.
%and we see that our data does not satisfy asymptotic scaling.
}
\label{fig:scaling_lambda}
\end{figure}

Looking at the ratio
$(1P-1S)/\Lambda_{\rm QCD}$ it is clear that our data does not satisfy the strict
criterion of asymptotic scaling, see Fig. \ref{fig:scaling_lambda}.
In this plot we determined $\Lambda_{\rm QCD}$ from the 2-loop formula 
in the $\overline{MS}$ scheme, 
\be
\Lambda_{\rm QCD} = 
\pi \left(\frac{\alpha b_0}{4\pi}\right)^{(-b_1/2b_0^2)} \exp\left(-\frac{2\pi}{b_0 \alpha}\right)
\left( 1 + \alpha \frac{b_1^2 - b_2b_0}{8\pi b_0^3}\right)~~, 
\ee
where the $\overline{MS}$ coupling constant 
$\alpha=\alpha_{\overline{MS}}(\pi/a)$ is 
estimated with a tadpole-corrected one-loop relation defined by
\be
\frac{1}{\alpha_{\overline{MS}}(\pi/a)} = 
\frac{(3.648 P-2.648 R)}{\alpha_0} + 4\pi (0.0589 + 0.0218 N_f)~~. 
\ee
Here $\alpha_0=g^2/4\pi$ denotes the bare coupling, and 
$P$ and $R$ are, respectively,  $1\times 1$ and $1\times 2$ Wilson loops 
normalized to unity for $U_\mu(x)=1$.  

\begin{figure}
\begin{center}
\leavevmode
\hbox{\epsfxsize = 12 cm \epsffile{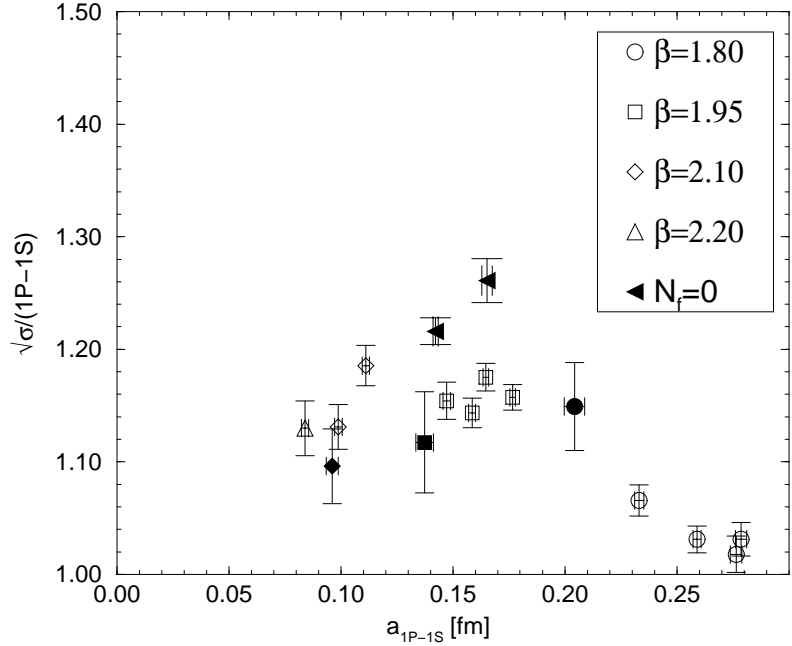}}
\end{center}
\caption{In contrast to Figs. \ref{fig:mrho_1P-1S} and \ref{fig:scaling_lambda}, we observe 
a better scaling for the ratio $\sqrt\sigma/(1P-1S)$ on our finer lattices.
%This indicates that discretisation errors already cancel in this ratio.
}
\label{fig:scaling_sigma}
\end{figure}

Within the effective approach of NRQCD, we cannot extrapolate such scaling violations
away and it is crucial to find other ratios in which the scaling violations
cancel each other already at finite lattice spacing. 
In Fig. \ref{fig:scaling_sigma} we show a test of this nature for the 
string tension, which shows a better scaling.  
Here we plot as open symbols the data obtained from 4 different sea quark
masses. At $\beta=2.20~(2.10)$ we only measured the lightest (and heaviest)
sea-quark mass, corresponding to $m_\pi/m_\rho \approx 0.60~(0.80)$.
This figure also suggests that the string tension, in units of the $1P-1S$ 
mass splitting, 
is smaller for 2-flavour QCD when compared to the quenched ($N_f=0$) theory.

\subsection{Hyperfine Splitting}

\begin{figure}
\begin{center}
\leavevmode
\hbox{\epsfxsize = 12 cm \epsffile{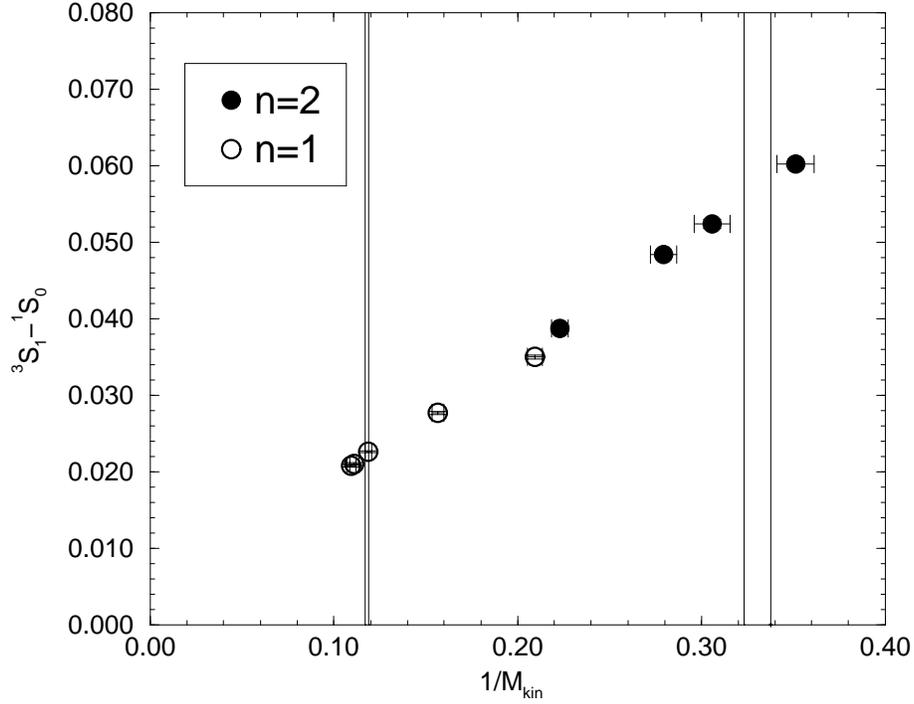}}
\end{center}
\caption{Mass dependence of hyperfine splitting. Here we show the
strong mass-dependence of the hyperfine splitting, ${^3}S_1-{^1}S_0$,
plotted against the inverse kinetic mass at $(\beta,\kappa)=(1.95,0.1375)$. 
The vertical lines denote the regions of the bottomonium and charmonium system.
This splitting is clearly very sensitive to the parameters of NRQCD.  
All data points are from updates with
$O(mv^6,a^2)$ and $n=1,2$ denotes different values of the stability parameter.}
\label{fig:hfs_vs_am}
\end{figure}

Quenching effects are also expected to show up in short-range quantities,
since they are particularly sensitive to the shape of the QCD potential.
In \cite{kaneko,sesam_coupling} this difference has been demonstrated explicitly by observing 
a change in the Coulomb coefficient of the static potential. 
In the context of heavy quarkonia, the hyperfine splitting is such a
UV-sensitive quantity which should be particularly susceptible to changes in
the number of flavours and the sea quark mass.
The prediction from potential models is
\be
^3S_1 - {^1S}_0 = \frac{32\pi\alpha_s(q)}{9m_Q^2}|\Psi(0)|^2~~.
\label{eq:hfs_pot}
\ee
In our study this is the most accurately measured quantity and 
it is clearly very sensitive to the value of the heavy quark mass, 
see Fig. \ref{fig:hfs_vs_am}.
As has been noticed previously, higher order relativistic and
radiative corrections are equally important for precision measurements of the hyperfine splitting
in bottomonium \cite{omv6,spitz} and even more so for charmonium \cite{trottier}.
Here we employ ${\cal O}(mv^6,a^2)$ as the standard accuracy throughout this paper.

\begin{figure}
\begin{center}
\leavevmode
\hbox{\epsfxsize = 12 cm \epsffile{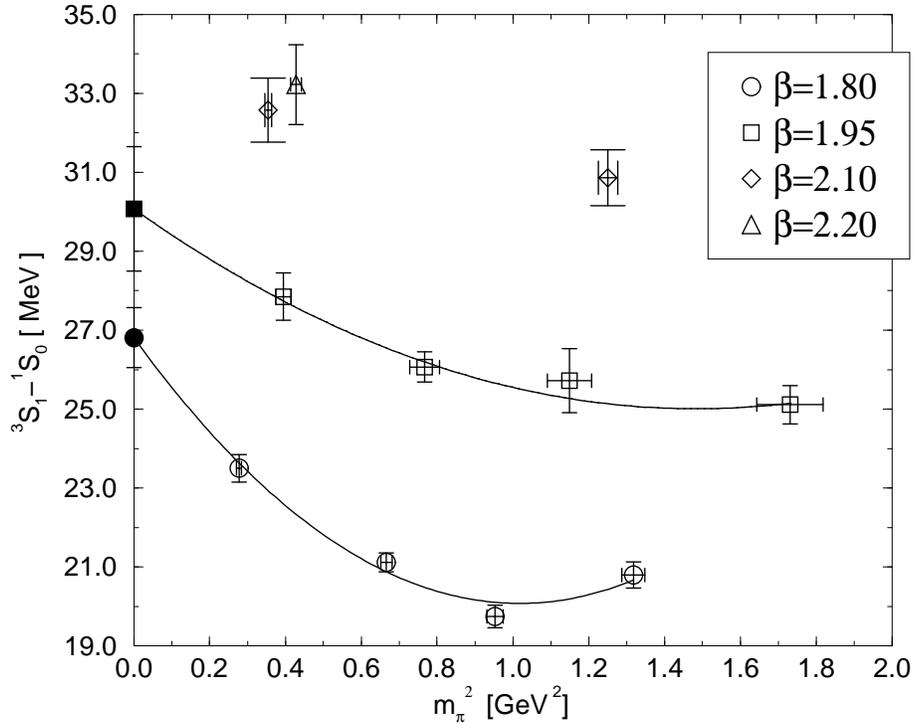}} 
\end{center}
\caption{Hyperfine splitting vs. $m_\pi^2$. Here we collect the data for the
the ${^3}S_1-{^1}S_0$ splitting in bottomonium from all values of
$(\beta,\kappa)$. A clear dependence on the sea quark mass can be seen.
The linear-plus-quadratic fit curves are shown as solid lines.
Here we used the $1P-1S$ splitting to determine the lattice spacing.}
\label{fig:hfs_vs_ampi2}
\end{figure}

The equation (\ref{eq:hfs_pot}) involves a direct dependence 
on both the strong coupling and the wavefunction at the origin, 
which makes the hyperfine splitting an ideal quantity to study 
quenching effects. 
% and other systematic errors. 
% old
% Indeed, we observe a clear rise of the hyperfine splitting 
% as we decrease the sea quark mass, see Fig. \ref{fig:hfs_vs_ampi2}.
% new 
Here we also observe a clear rise of the hyperfine splitting 
as we decrease the sea quark mass, see Fig. \ref{fig:hfs_vs_ampi2}.

\begin{figure}
\begin{center}
\leavevmode
\begin{tabular}{c}
\hbox{\epsfxsize = 12 cm \epsffile{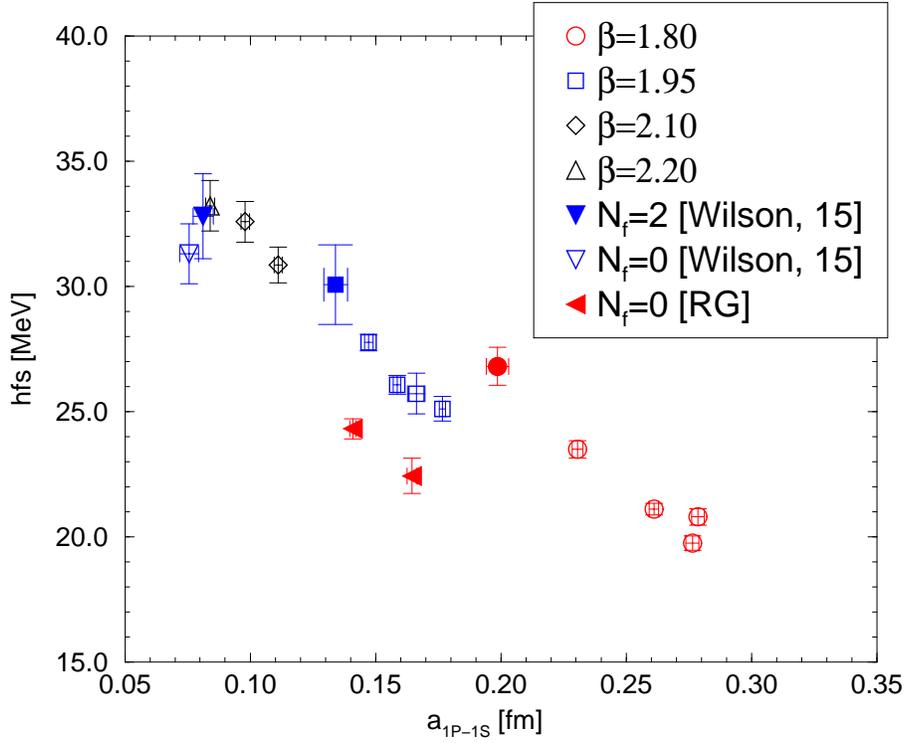}}
\end{tabular}
\end{center}
\caption{Scaling violations for hyperfine splitting. Open symbols
correspond to runs with different sea quark mass. Filled symbols
denote the dynamical data after chiral extrapolation and results 
with $N_f=0$. We used $1P-1S$ splitting to determine the lattice spacing.}
\label{fig:hfs_scaling}
\end{figure}

In Fig. \ref{fig:hfs_scaling} we collected all our dynamical results for the
hyperfine splitting over the range of $0.1-0.2$ fm. 
Here we plotted the data from each sea-quark mass as open
symbols and used the experimental $1P-1S$ splitting to convert lattice data
into MeV.  One should keep in mind that these
points correspond to unphysical bottomonium in a world of different sea quark
masses. 
% NEW
We also plot as filled symbols the results of our naive chiral extrapolation
as described in the previous section.
 
%The physical points obtained by chiral extrapolations of the sea 
%quark mass are plotted with filled symbols.

At around 0.10 fm we notice a very good agreement with the only previous
calculation \cite{spitz}. These authors have performed a dynamical simulation
at a single lattice spacing using Wilson glue and unimproved sea-quarks.
For the bottom quarks they used an NRQCD formulation with the same accuracy, 
$O(mv^6,a^2)$, as in this study.

An unpleasant feature with our results in Fig. \ref{fig:hfs_scaling} 
is lack of scaling; for both the full and quenched case we find 
scaling violations of about 100 MeV/fm for the hyperfine splitting. 
Nonetheless, 
we do find several indications for sea quark effects in our results. 
First we notice that, if it were not for
sea quark effects, then all points in Fig. \ref{fig:hfs_scaling} would lie on a
universal curve which is not the case. 
This is a strong indication that for this quantity we have to expect effects
%Moreover, there is a clear gap between
%our quenched and dynamical data at the same lattice spacing. This is a very
%strong indication that we have to expect effects 
of the order of 3-5 MeV when going from zero to two flavour QCD.

\begin{figure}
\begin{center}
\leavevmode
\hbox{\epsfxsize = 12 cm \epsffile{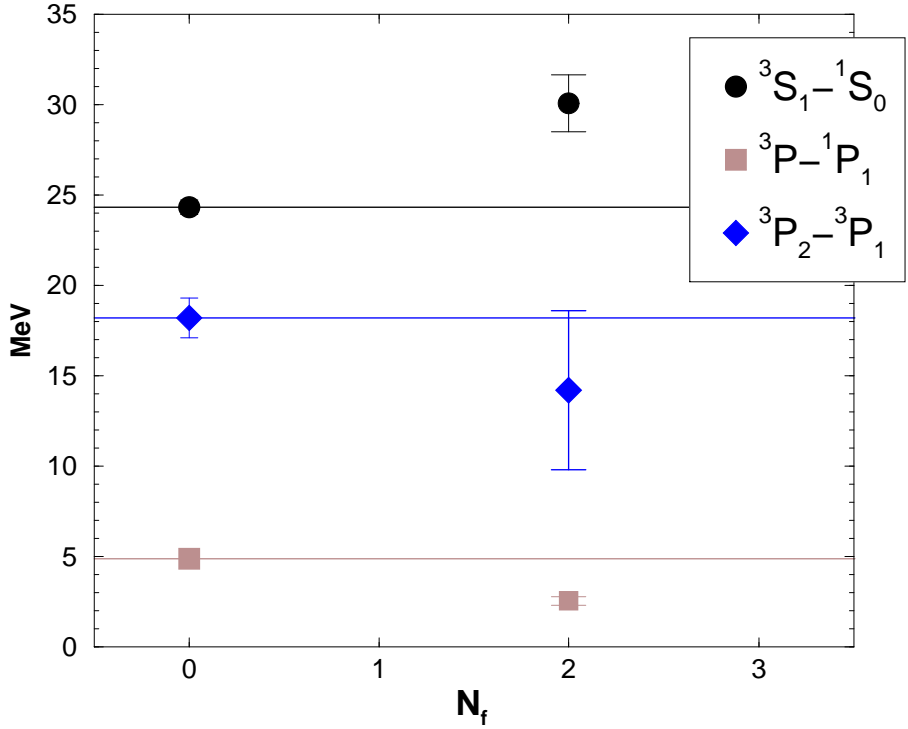}}
\end{center}
\caption{Direct comparison of the bottomonium spin structure for quenched 
and full QCD at the same lattice spacing 
of $a \approx 0.14$ fm. The $N_f=2$ data is taken from the chiral
limit of our measurements at $\beta=1.95$.}
\label{fig:direct_hfs}
\end{figure}

To substantiate this observation 
%In Fig. \ref{fig:direct_hfs} 
we make a direct comparison of 
quenched and dynamical calculations at the {\it same} lattice spacing 
of 0.14 fm in Fig. \ref{fig:direct_hfs}.  
For the ${^3}S_1-{^1}S_0$ splitting replotted from Fig. \ref{fig:hfs_scaling}, 
a clear increase of around 5 MeV (20\%) 
represents more than a $5\sigma$-effect, which reflects the accuracy in our
determination of this quantity.  
On the other hand, the hyperfine splitting in $P$-states is reduced as we approach a
more realistic description of QCD. Within potential models, states with $L \ne
0$ are not sensitive to the contact term of the spin-spin interaction. However
the perturbative expression for a higher order radiative corrections
\cite{phfs} gives a ${^3}P-{^1}P_1$ splitting opposite in sign to our values. 
Experimentally, the spin-triplet states are well established, 
but ${^1}S_0$ and ${^1}P_1$ have yet to be confirmed for bottomonium.

We comment that our data for charmonium (Tab. \ref{tab:charm_beta180}) also 
indicates a rise in the hyperfine splitting towards the chiral limit. 
It is, however, apparent that such a rise can not explain
the discrepancy between the NRQCD predictions and the experimentally
observed spin structure. We confirm an earlier observation \cite{trottier} that the velocity expansion
is not well controlled for charmonium where $v_c^2 \approx 0.3$.

\subsection{Fine Structure}
\label{sec:fs}
In the continuum, the fine structure in quarkonia is due to the different ways
in which the spin can couple to the orbital angular momentum of the bound
state. In our approach, the spin-orbit term and the tensor term of potential models can be
traced back to the $c_3$ term in (\ref{eq:dh}).
A correct description of the fine structure will therefore 
require a proper determination of $c_3$ and the corrections to this term.

On the lattice we have also additional splittings with no continuum analogue.
For example, the ${^3}P_{2E}-{^3}P_{2T}$ splitting is known to be a pure discretisation error
since the lattice breaks the rotational invariance of the continuum and causes the
$J=2$ tensor to split into two irreducible representations of the orthogonal
group: $T_2$ and $E$. Indeed, for both the dynamical calculations as well as
the quenched data, we observe a significant reduction of this splitting when the
lattice spacing is decreased; the splitting diminishes 
from about 1.5 MeV at $a\approx 0.2$ fm to 
0.5 MeV at $a\approx 0.1$ fm for the dynamical case.

\begin{figure}
\begin{center}
\leavevmode
\hbox{\epsfxsize = 12 cm \epsffile{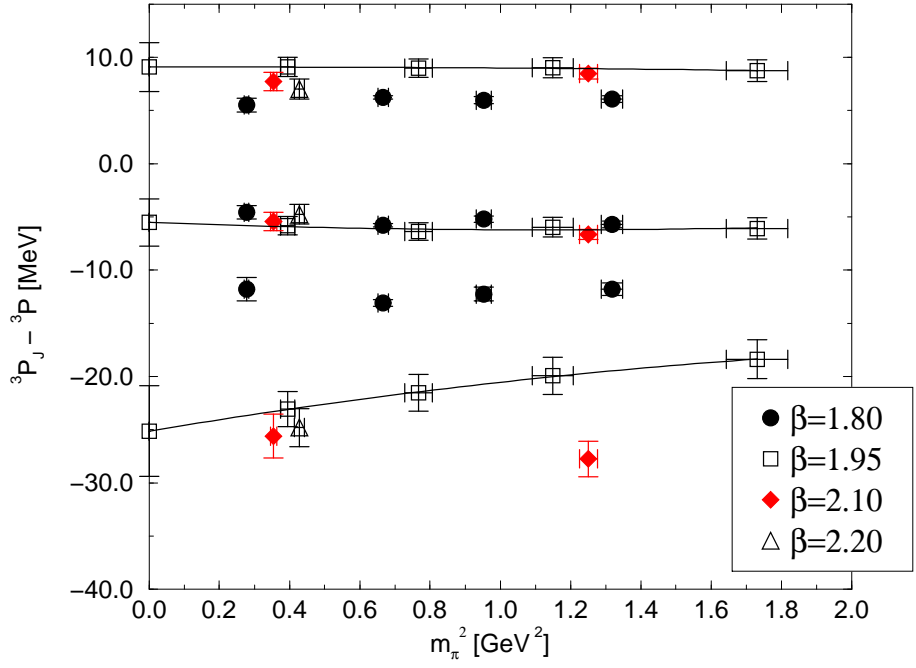}}
\end{center}
\caption{Fine structure in Bottomonium. Here we plot (top to
bottom) ${^3}P_2,{^3}P_1$ and ${^3}P_0$ relative to the spin averaged triplet state: ${^3}P \equiv 1/9 
(5 {^3}P_2 + 3 {^3}P_1 + 1 {^3}P_0)$. The corresponding experimental values
are: 13 MeV, -8 MeV and -40 MeV. }
\label{fig:fs}
\end{figure}

In Fig. \ref{fig:fs} we show our results for the fine structure. 
For ${^3}P_2$ and ${^3}P_1$  we observe no clear dependence on 
the sea quark mass.
This is not totally unexpected since $P$-state wavefunctions vanish at the
origin and should not be as strongly dependent on changes in the UV-physics.
In any case such small changes would be difficult to resolve within our
statistical errors. 

From Fig. \ref{fig:fs} we can also
see a better scaling behaviour of the $P$ states, apart perhaps from the
${^3}P_0$, where scaling violations still obscure the chiral behaviour. 
The latter has $J=0$ and therefore we may expect that for this
state restoration of rotational invariance is particularly important.

On our finer lattices we observe an increase of the ${^3}P_0-{^3}P$
splitting, closer towards the experimental value of $-40$ MeV.
We take this as an indication that a better control of the lattice spacing
errors and radiative corrections is necessary to reproduce this quantity in
NRQCD lattice calculations.
The other splittings, ${^3}P_2-{^3}P$ and ${^3}P_1-{^3}P$, deviate only by a few MeV
from their experimental values of 12 MeV and $-8$ MeV, which could be due to 
missing dynamical flavours ($N_f=3$), higher order relativistic effects and radiative corrections.
%The predictions for the other splittings, ${^3}P_2-{^3}P$ and ${^3}P_1-{^3}P$, are
%already very close to their experimental values of 12 MeV and $-8$ MeV. 
%The remaining deviations of
%only a few MeV could be due to missing dynamical flavours ($N_f=3$), higher
%order relativistic effects and radiative corrections.

\begin{figure}
\begin{center}
\leavevmode
\hbox{\epsfxsize = 12 cm \epsffile{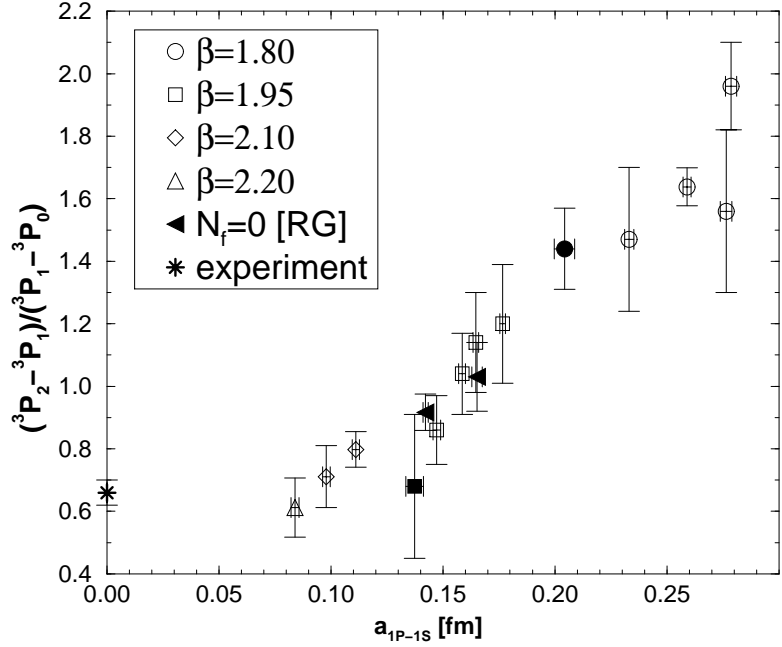}}
\end{center}
\caption{Fine structure ratio in Bottomonium. Here the lattice data should 
be compared to the experimental value of 0.66(4). It is apparent that there
are still large underlying scaling violations, but no clear sea quark dependence.}
\label{fig:Rfs}
\end{figure}

We take the fine structure ratio,
$R_{fs}=({^3}P_2-{^3}P_1)/({^3}P_1-{^3}P_0)$, as a particularly sensitive
quantity to measure the internal consistency of the $P$-triplet structure.
This quantity should be less sensitive to radiative corrections of the NRQCD
coefficients away from their tree-level values. 
Previous NRQCD calculations had measured this quantity to be
much larger than 1, compared to the experimental value of 0.66(4).
We believe that this discrepancy is due to lattice spacing artefacts
as it is very sensitive to the implementation of ${\cal O}(a^2)$
improvement in the NRQCD formalism.
It is encouraging to see that this value is further reduced on our 
finer lattices, see Fig. \ref{fig:Rfs}. Notably, we do not observe 
any difference between our dynamical results and the quenched data.

\subsection{$2S-1S$ Splitting}
Another spectroscopic quantity which has attracted much attention
is the $2S-1S$ splitting, since it should also be sensitive to the short-range
potential.
On conventional lattices such higher excitations are difficult to resolve and 
requires delicate tuning to minimise the mixing of the $2S$ with the ground
state. Given our rather coarse lattices we did not attempt to perform a
systematic study of this quantity, but in the context of this section it is
important to notice that we do not observe any chiral dependence of the
ratio, $R_{2S}=(2S-1S)/(1P-1S)$.
 
\begin{figure}
\begin{center}
\leavevmode
\hbox{\epsfxsize = 12 cm \epsffile{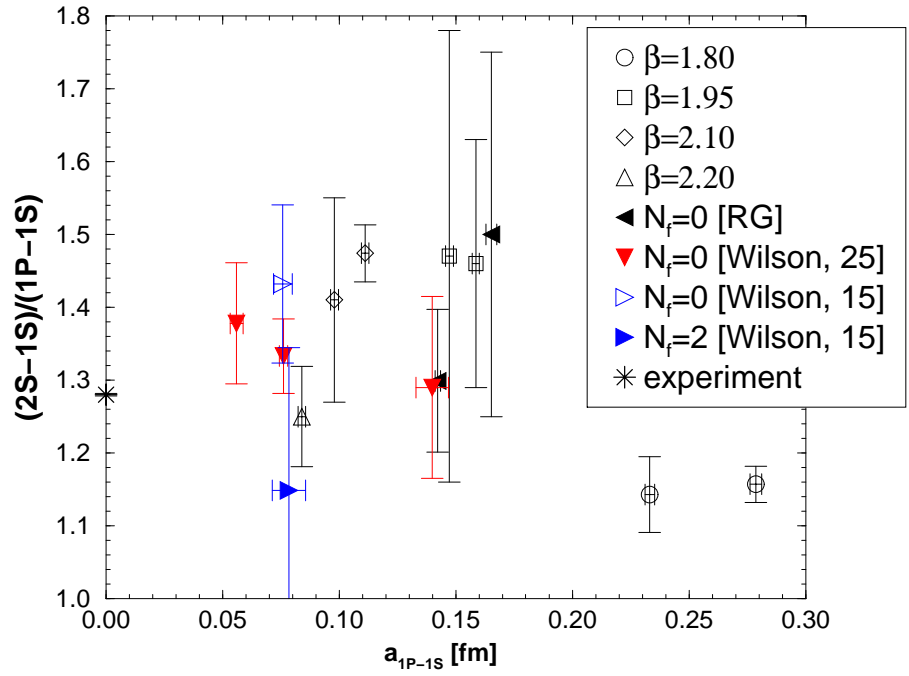}}
\end{center}
\caption{Here we show a scaling plot of the ratio
$R_{2S}=(2S-1S)/(1P-1S)$. It is apparent that one needs much smaller
statistical errors to resolve any systematic effects.
We show our results from different sea quark masses along with
representative results from other collaborations.}
\label{fig:R2S}
\end{figure}

In Fig. \ref{fig:R2S} we compiled representative data from other groups 
\cite{spitz,scaling} along with our new results from the RG-action. 
Within the large errors we cannot resolve a discrepancy between the
experiment and the lattice data.
This result is in contrast to the previous determinations of this quantity 
which claim to
see deviations due to missing sea quarks \cite{spitz,scaling}.

Observing such deviations is certainly plausible as this ratio is thought to be sensitive to differences in the underlying short-range potential.
However, for the same reason we should also expect large scaling violations.
Interestingly, on our coarsest dynamical lattices we even observe smaller
values of $R_{2S}$,  which we take as an indication of large discretisation errors. 
Apart from this very coarse lattice data, we cannot resolve either scaling violations
or quenching effects. We feel that it requires a much better resolution of the higher excited
states, which is hard to achieve on isotropic lattices.
Future lattice studies will need optimised meson operators or finer
temporal discretisations to observe these effects.

\section{Conclusion}
\label{sec:conclusion}
We have demonstrated that dynamical sea quarks have a significant effect
on the spectrum of heavy quarkonia. Namely the hyperfine splitting,
${^3}S_1-{^1}S_0$, 
% NEW
is raised by almost $20\%$ when going from zero to two dynamical flavours.
%shows a clear trend upwards when the ratio $m_\pi/m_\rho$ is lowered.
The efficiency of the NRQCD approach has played an important role
to establish such effects, but the numerical simplicity of this approach is 
offset by additional systematic errors, which have to be controlled.
The sensitivity of the spin-structure to relativistic, ${\cal O}(mv^6)$, and radiative,
${\cal O}(\alpha)$, corrections was well known before we started this work.
Here we demonstrated that quenching errors are equally important for
precision measurements of the spectrum of heavy quarkonia.
Perhaps more worrying are scaling violations, which we could resolve in many
quantities. Without a proper control of lattice spacing artefacts it is not
possible to make predictions for such UV-sensitive quantities as the hyperfine 
splitting on the lattices we used here.

While the lattice predictions for ${^3}P_1-{^3}P$ and ${^3}P_2-{^3}P$ agree
well with their experimental values,
the determination of the ${^3}P_0$ is more problematic and we still observed large
deviations from the experimental value when the $1P-1S$ splitting is used to
set the scale.
Clearly much work remains to be done to reduce the systematic errors in 
heavy quark physics to the same degree as the statistical ones.
We feel that this may be difficult to achieve within the non-relativistic 
framework.  A better description of the NRQCD coefficients or a
relativistic treatment is in order to describe the spin structure in quarkonia accurately. 
From this work it is apparent that full QCD simulations are also necessary to achieve such a goal.

For less accurate quantities such as $2S-1S$ it is more difficult
to reach a similar conclusion and we leave a decisive observation
of both scaling violations and sea quark effects to future studies with
refined methods.

\begin{center}{\bf Acknowledgments}\end{center}
The calculations were done using workstations and the CP-PACS facilities at
at the Center for Computational Physics at the University of Tsukuba.
This work is supported in part by the Grants-in-Aid of Ministry of
Education (Nos. 09304029, 10640246, 10640248, 10740107, 11640250, 11640294,
11740162).  TM and AAK are supported by the 
JSPS Research for the Future Program (Project No. JSPS-RFTF 97P01102). 
SE, KN and HPS are JSPS Research Fellows.

\newpage

\begin{table}[t]
\begin{tabular}{ccccccc}
\hline
\multicolumn{7}{c}{dynamical simulation, $N_f=2$}\\
\hline
$\beta,~(L^3 \times T)$ & $c_{SW}$ & $\kappa$ & $m_Q$              & $u_{0P}$       & $u_{0L}$      & traj./cfgs.   \\
\hline                                
1.8, ($12^3\times 24$)  & 1.60     & 0.1409   & 2.20, 4.00, 5.30, 6.00, 6.10, 7.00   &  0.836885(13)  & 0.77164(12)  & 6250/640   \\
                        &          & 0.1430   & 2.10, 5.20, 5.50, 5.85         &  0.838807(17)  & 0.77584(15)  & 5000/512    \\
                        &          & 0.1445   & 2.06, 5.61, 5.80               &  0.840627(16)  & 0.77994(16)  & 7000/360    \\
                        &          & 0.1464   & 1.77, 1.90, 2.00, 5.00, 5.18, 5.50   &  0.843909(24)  & 0.78770(18)  & 5250/408     \\
\hline                                
1.95,  ($16^3\times 32$)& 1.53     & 0.1375   & 1.20, 1.38, 1.50, 2.00, 4.00, 4.29  & 0.8624838(78)  & 0.817086(41) & 7000/120    \\
                        &          & 0.1390   & 1.22, 1.40, 3.80, 4.00              &  0.8637715(81) & 0.820962(46) & 7000/256    \\
                        &          & 0.1400   & 1.19, 1.25, 3.40, 3.60, 3.70        &  0.8647304(81) & 0.824140(28) & 7000/400   \\
                        &          & 0.1410   & 1.06, 1.15, 3.30, 3.40              &  0.865788(10)  & 0.827498(31) & 5000/400 \\
\hline                                
2.10, ($24^3\times 48$) & 1.47     & 0.1357   & 2.45, 2.72                     &  0.8793870(40) & 0.850170(25) & 2000/192     \\
%                        &          & 0.1367   & --                                   &  0.8798149(37) & 0.851953(31)&  2000/ ---  \\
%                        &          & 0.1374   & --                                   &  0.8801334(37) & 0.853046(20)&  2000/ ---  \\
                        &          & 0.1382   & 1.95, 2.24, 2.65               &  0.8805090(44) & 0.854604(20)&  2000/192     \\
\hline                                
%2.20, ($24^3\times 48$) & 1.44     & 0.1351   & --                                   &  0.8873657(36) & 0.864203(30)&  2000/---       \\
%                        &          & 0.1358   & --                                   &  0.8875764(25) & 0.864962(21)&  2000/---        \\
%                        &          & 0.1363   & --                                   &  0.8877262(29) & 0.865752(24)&  2000/---       \\ 
2.20, ($24^3\times 48$)  & 1.44     & 0.1368   & 1.95,  2.21                    &  0.8878887(29) & 0.866357(23)&  2000/128     \\
\hline                        
\multicolumn{7}{c}{quenched simulation, $N_f=0$ }  \\
\hline                        
\multicolumn{3}{c}{ $\beta,~(L^3 \times T)$ }            & 
\multicolumn{1}{c}{ $m_Q$}     & 
\multicolumn{1}{c} {$u_{0P}$ }   & 
\multicolumn{1}{c}{$u_{0L}$ }       &   
\multicolumn{1}{c}{updates/cfgs. }   \\
\hline                                
\multicolumn{3}{c}{2.187, ($16^3\times 32$) }             & 
\multicolumn{1}{c}{3.70,  4.00}   &  
\multicolumn{1}{c}{0.8772362(22)} & 
\multicolumn{1}{c}{0.831789(55)}   &  
\multicolumn{1}{c}{20000/200} \\
\multicolumn{3}{c}{2.281, ($16^3\times 32$)}  &
\multicolumn{1}{c}{3.20,  3.40}   &  
\multicolumn{1}{c}{0.8855537(18)} & 
\multicolumn{1}{c}{0.847829(41)} &
\multicolumn{1}{c}{20000/200}\\
\hline
\end{tabular}
\caption{Simulation parameters for this study. The last column gives the
number of configurations employed for this work.  The quenched runs 
are made at $\beta=2.187$ and 2.281 so that the string tension 
matches with those of the $N_f=2$ runs at 
$(\beta,\kappa)=(1.95, 0.1390)$ and at $(1.95, \kappa_c)$ on a $16^3\times 32$ 
lattice.}
\label{tab:simulation}
\end{table}

\begin{table}[htp]
\begin{tabular}{lll||ll}
\hline
$\beta$                        & 2.187       &  2.187     & 2.281        & 2.281         \\
$M_b$                          & 3.70        &  4.00      & 3.20         & 3.40           \\
$M_{kin}$ [GeV]                & 9.04(23)    &  9.95(27)  & 9.110(94)    & 9.77(10)      \\
$a(1P-1S)$ [fm]                & 0.1653(23)  & 0.1637(15) & 0.1423(12)   & 0.1400(12)       \\
\hline                                      
$R_{2S}$                       &  1.50(25)   &  1.65(59)  & 1.299(98)    & 1.26(11)       \\
%$R_{2P}$                      &  4.5(2.3)   &  --        &   --         & --             \\
\hline                                      
${^3}S_1-{^1}S_0$ [MeV]        & 23.12(34)   &  21.68(22) &  24.88(25)   & 23.91(26)      \\
%(2) ${^3}S_1-{^1}S_0$ [MeV]   & 29(187)     &  --        &  20(65)      & 22(70)        \\
\hline                                      
${^3}P -{^1}P_1$ [MeV]         &  4.03(61)   &   3.79(61) &  4.97(35)    &  4.87(36)        \\
${^3}P -{^3}P_0$ [MeV]         & 20.4(1.3)   &  19.5(1.4) &  29.8(1.0)   & 28.1(1.2)   \\
${^3}P -{^3}P_1$ [MeV]         &  6.73(65)   &   6.46(74) &   9.07(58)   &  8.93(59)    \\
${^3}P_2 -{^3}P$ [MeV]         &  6.83(56)   &   6.44(65) &   9.26(54)   &  9.23(51)  \\
${^3}P_{2E} -{^3}P_{2T}$ [MeV] &  1.67(45)   &   1.43(28) &   0.91(36)   &  0.87(35)  \\
\hline                                      
${^3}P_{2} -{^3}P_{1}$ [MeV]   & 13.6(1.2)   &  12.9(1.4) &   18.8(1.1)  & 18.2(1.1)   \\
${^3}P_{1} -{^3}P_{0}$ [MeV]   & 13.22(76)   &  12.73(82) &   20.53(62)  & 19.39(65) \\
$R_{fs}$                       &  1.03(11)   &   1.01(13) &   0.917(58)  &  0.937(64)  \\
%$R_{ss/fs}$                    &  1.75(10)   &   1.70(11) &   1.212(39)  &  1.233(43)  \\
\hline
\end{tabular}
\caption{Bottomonium spectrum from quenched run at $\beta=2.187$ and $\beta=2.281$.
These results should be compared directly to the $N_f=2$ data at the similar lattice
spacing: $(\beta,\kappa)=(1.95,0.1390)$ and $(1.95,\kappa_c)$, respectively.
We also illustrate the effects of little changes in the quark mass on the spectrum. The difference for
the hyperfine splitting, ${^3}S_1-{^1}S_0$, is most noticeable. For the
other splittings we see indications of such a suppression as the mass is
increased, but it is much less pronounced within the errors. 
Scaling violations can be observed in ${^3}P_{2E}-{^3}P_{2T}$, as discussed in
the main text. All the other splittings are suppressed on the coarser lattice.
On the finer lattice, the ratio $R_{fs}=({^3}P_2-{^3}P_1)/({^3}P_1-{^3}P_0)$ is
closer to its experimental value: 0.66(4).}
\label{tab:quenched}
\end{table}

\begin{table}[htp]
\begin{tabular}{llllll}
\hline
$\kappa$                       & 0.1409     & 0.1430     & 0.1445     & 0.1464     & $\kappa_c$ \\
$m_\pi/m_\rho$                 & 0.80599(75)& 0.7531(13) & 0.6959(20) & 0.5480(45) &             \\
$M_b$                          & 5.87(5)    & 5.85       & 5.61       & 5.10(5)    &             \\
$M_{kin}$ [GeV]                & 9.46(12)   & 9.43(10)   & 9.530(80)  & 9.46(11)   &             \\
$a(1P-1S)$ [fm]                & 0.2787(25) & 0.2765(26) & 0.2611(19) & 0.2306(21) & 0.1987(32)  \\
$a(\rho_{PQ})$ [fm]            & 0.2622(11) & 0.2560(16) & 0.2462(13) & 0.2246(18) & 0.2040(40)  \\
\hline
$R_{2S}$                       &  1.157(25) &            &            &  1.143(52) &             \\
%$R_{2P}$                      &  1.89(14)  &            &            &  --        &             \\
\hline
${^3}S_1-{^1}S_0$ [MeV]        & 20.80(33)  &  19.75(29) &  21.11(24) & 23.50(35)  & 26.81(76)   \\
%(2) ${^3}S_1-{^1}S_0$ [MeV]   & 12.7(9.8)  &            &            & 10(33)     &             \\
\hline
${^3}P -{^1}P_1$ [MeV]         &  3.75(24)  &  3.91(18)  &   3.64(54) &  3.40(56)  &  2.82(1.03)           \\
${^3}P -{^3}P_0$ [MeV]         & 11.80(59)  & 12.26(65)  &  13.09(33) & 11.8(1.1)  & 13.61(75)   \\
${^3}P -{^3}P_1$ [MeV]         &  5.71(33)  &  5.20(30)  &   5.78(15) &  4.57(64)  &  5.64(12)   \\
${^3}P_2 -{^3}P$ [MeV]         &  6.08(31)  &  5.97(33)  &   6.23(16) &  5.51(64)  &  6.12(39)   \\
${^3}P_{2E} -{^3}P_{2T}$ [MeV] &  1.84(23)  &  1.56(26)  &   1.67(13) &  1.58(47)  &  1.47(31)  \\
\hline
${^3}P_{2} -{^3}P_{1}$ [MeV]   & 11.81(63)  & 11.18(60)  &  12.02(30) & 10.1(1.3)  &  11.69(77)            \\
${^3}P_{1} -{^3}P_{0}$ [MeV]   &  6.02(31)  &  7.19(42)  &   7.33(23) &  6.91(63)  &  8.13(46)           \\
$R_{fs}$                       &  1.96(14)  &  1.55(12)  &  1.640(66) &  1.46(23)  &  1.44(13)  \\
%$R_{ss/fs}$                   &  3.46(18)  &  2.75(15)  &  2.879(90) &  3.40(31) &     \\
\hline
\end{tabular}
\caption{Bottomonium results from $\beta=1.80$. An error in the quark mass parameter
indicates that we have interpolated to this value in order to reproduce the
Bottomonium at any given $(\beta,\kappa)$. Where this error is not given we
accepted the value of the tuned quark mass. 
For the hyperfine splitting we could fit the data to a linear-plus-quadratic
dependence in the quark mass, but for the fine structure the quadratic part was
ill-determined and we resorted to linear or constant fits if their Q-value was acceptable.}
\label{tab:ups_beta180}
\end{table}

\begin{table}[htp]
\begin{tabular}{llllll}
\hline
$\kappa$                       & 0.1375     & 0.1390     & 0.1400    & 0.1410     & $\kappa_c$ \\
$m_\pi/m_\rho$                 & 0.80484(89)& 0.7514(14) & 0.6884(15)& 0.5862(33) &             \\
$M_b$                          & 4.00       & 3.80       &  3.70     & 3.40       &             \\
$M_{kin}$ [GeV]                & 9.40(16)   & 9.43(22)   & 9.43(10)  & 9.49(17)   &             \\
$a(1P-1S)$ [fm]                & 0.1767(13) & 0.1662(35) &0.1586(15) & 0.1473(17) & 0.1341(48)   \\
$a(\rho_{PQ})$ [fm]            & 0.1974(11) & 0.1860(12) &0.1791(10) & 0.1625(13) & 0.1451(33)  \\
\hline
$R_{2S}$                       &  --        &  1.242(72) & 1.46(17)  & 1.47(31)   &             \\
%$R_{2P}$                      &  --        &            &  --       & --         &             \\
\hline
${^3}S_1-{^1}S_0$ [MeV]        &  25.11(49) &  25.72(81) & 26.07(38) & 27.85(60)  &  30.07(1.58)  \\
%(2) ${^3}S_1-{^1}S_0$ [MeV]   &  --        &            & 45(120)   & --         &             \\
\hline
${^3}P -{^1}P_1$ [MeV]         &  2.41(66)  &   2.26(63) & 2.50(64)  &  2.67(32)  &  2.55(24)   \\
${^3}P -{^3}P_0$ [MeV]         & 18.4(1.8)  &  20.0(1.8) &21.5(1.7)  & 23.1(1.7)  & 24.9(5.5)    \\
${^3}P -{^3}P_1$ [MeV]         &  6.08(99)  &   5.97(93) & 6.38(82)  &  5.83(86)  &  5.5(2.2)   \\
${^3}P_2 -{^3}P$ [MeV]         &  8.7(1.0)  &   9.01(93) & 8.98(85)  &  9.10(89)  &  9.0(2.3)    \\
${^3}P_{2E} -{^3}P_{2T}$ [MeV] &  2.05(15)  &   1.50(13) & 1.41(20)  &  1.22(74)  &  1.33(29)    \\
\hline
${^3}P_{2} -{^3}P_{1}$ [MeV]   & 14.8(2.0)  &  15.1(1.8) & 15.5(1.6) & 14.8(1.7)  &  14.2(4.4)  \\
${^3}P_{1} -{^3}P_{0}$ [MeV]   & 12.4(1.1)  &  13.2(1.0) & 14.9(1.1) & 17.4(1.0)  &  20.9(2.5)  \\
$R_{fs}$                       &  1.19(19)  &  1.14(16)  & 1.04(13)  & 0.85(11)   &   0.68(23)  \\
%$R_{ss/fs}$                   &  2.02(18)  &  1.97(16)  & 1.76(13)  & 1.604(93)  &             \\
\hline
\end{tabular}
\caption{Bottomonium results from $\beta=1.95$.}
\label{tab:beta195}
\end{table}

\begin{table}[htp]
\begin{tabular}{lllll}
\hline
$(\beta,\kappa)$               & (2.10,0.1357)   & (2.10,0.1382) & (2.20,0.1368))  & experiment    \\
$m_\pi/m_\rho$                 & 0.8066(16)      & 0.5735(48)    & 0.6320(70)      & 0.18          \\
$M_b$                          & 2.45            & 2.24          & 1.95            & --              \\
$M_{kin}$ [GeV]                & 9.34(16)        & 9.58(17)      & 9.46(20)        & 9.46037(21)            \\
$a(1P-1S)$ [fm]                & 0.1112(16)      & 0.0980(17)    & 0.0840(18)      & --         \\
$a(\rho_{PQ})$ [fm]            & 0.1361(15)      & 0.1169(17)    & 0.0946(16)      & --       \\
\hline                                                                  
$R_{2S}$                       &  1.474(39)      & 1.41(14)      &  1.250(69)      &  1.2802(15)         \\
%$R_{2P}$                      &  2.28(16)       & 2.33(18)      &  2.46(21)       &  1.8185(20)         \\
\hline                                                                  
${^3}S_1-{^1}S_0$ [MeV]        & 30.86(71)       & 32.58(81)     &  33.2(1.0)      & --                \\
%(2) ${^3}S_1-{^1}S_0$ [MeV]   & 41.0(3.1)       & ---           &  77(67)         & --          \\
\hline                                                                  
${^3}P -{^1}P_1$ [MeV]         &  2.08(63)       &  1.58(47)     &  2.24(20)       & --           \\
${^3}P -{^3}P_0$ [MeV]         & 27.7(1.7)       & 25.60(2.1)    & 24.8(1.8)       & 40.3(1.4)           \\
${^3}P -{^3}P_1$ [MeV]         &  6.64(48)       &  5.42(87)     &  4.75(92)       &  8.2(8)           \\
${^3}P_2 -{^3}P$ [MeV]         &  8.46(50)       &  7.72(86)     &  7.00(94)       & 13.1(7)          \\
${^3}P_{2E} -{^3}P_{2T}$ [MeV] &  0.31(25)       &  0.48(19)     &  0.78(10)       & --           \\
\hline                                                                  
${^3}P_{2} -{^3}P_{1}$ [MeV]   & 15.20(97)       & 13.4(1.7)     &  12.0(1.8)      & 21.3(9)           \\
${^3}P_{1} -{^3}P_{0}$ [MeV]   & 19.07(89)       & 18.8(1.3)     &  19.7(1.2)      & 32.1(1.5)           \\
$R_{fs}$                       & 0.797(63)       & 0.713(99)     &   0.609(99)     & 0.66(4)           \\
%$R_{ss/fs}$                   & 1.619(74)       & 1.73(12)      &   1.69(10)      & --            \\
\hline
\end{tabular}
\caption{Bottomonium results from $\beta=2.10$ and 2.20.}
\label{tab:beta210}
\end{table}

\begin{table}[htp]
\begin{tabular}{lllllll}
\hline
$(\beta,\kappa)$               & (1.80,0.1409) & (1.80,0.1430) & (1.80,0.1445) & (1.80,0.1464)  & (1.95,0.1375) & exp. \\
$m_\pi/m_\rho$                 & 0.80599(75)   & 0.7531(13)    & 0.6959(20)    & 0.5480(45)     & 0.80484(89)   & 0.18        \\
$M_b$                          & 2.20          & 2.10          &  2.06         & 1.77           & 1.30(5)       &             \\
$M_{kin}$ [GeV]                & 3.019(87)     & 3.323(34)     &  3.589(46)    &  3.401(85)     & 3.01(12)      & 3.09688(4)            \\
$a(1P-1S)$ [fm]                & 0.2874(11)    & 0.2758(14)    & 0.2571(26)    & 0.2388(53)     & 0.1983(43)    &    \\
$a(\rho_{PQ})$ [fm]            & 0.2622(11)    & 0.2560(16)    & 0.2462(13)    & 0.2246(18)     & 0.1974(11)    &   \\
\hline                               
$R_{2S}$                       &  1.378(60)    & 1.29(10)      & 1.557(95)     &   2.02(34)     &  --           & 1.3009(31)   \\
%$R_{2P}$                      &  --           & 2.19(18   )   & --            &  --            &  --           &       --      \\
\hline                               
 ${^3}S_1-{^1}S_0$ [MeV]       & 49.60(35)     &  53.17(38)    & 54.02(67)     & 56.04(70)      &  55.5(2.8)    &  117(2)  \\
%(2) ${^3}S_1-{^1}S_0$ [MeV]   & 90(84)        &  78(132)      & 15(90)        & 165(237)       &  --           &   92(5)  \\
\hline                               
${^3}P -{^1}P_1$ [MeV]         &  3.66(23)     &  2.86(86)     & 3.25(41)      & 1.71(55)       &   4.0(2.0)    &  -0.86(25)          \\
${^3}P -{^3}P_0$ [MeV]         & 26.48(54)     & 31.52(74)     & 26.6(3.2)     & 31.4(2.1)      &   38.8(4.1)   &  110.2(1.0) \\
${^3}P -{^3}P_1$ [MeV]         &  5.14(32)     &  7.17(47)     & 2.6(1.7)      &  4.27(75)      &   0.90(1.0)   &   14.75(18) \\
${^3}P_2 -{^3}P$ [MeV]         &  8.86(79)     & 10.85(49)     & 7.9(1.6)      & 10.54(64)      &   7.2(1.0)    &   30.89(18)  \\
${^3}P_{2E} -{^3}P_{2T}$ [MeV] &  2.45(18)     &  3.00(46)     & 2.06(30)      &  2.18(66)      &   1.50(78)    &   --\\
\hline                               
${^3}P_{2} -{^3}P_{1}$ [MeV]   &  13.12(49)    &  18.15(76)    &  10.4(3.2)    & 14.8(1.3)      &   9.3(4.0)    &  45.64(18)  \\
${^3}P_{1} -{^3}P_{0}$ [MeV]   &  21.17(33)    &  24.40(49)    &  24.23(99)    & 29.25(84)      &  34.8(5.0)    &  95.4(1.0)  \\
$R_{fs}$                       &  0.620(25)    &  0.744(34)    &   0.43(13)    &  0.507(47)     &   0.27(12)    &  0.4783(54) \\
%$R_{ss/fs}$                   &  2.343(40)    &  2.179(46)    &  2.223(95)    & 1.916(60)      &   1.59(24)    &    \\
\hline
\end{tabular}
\caption{Charmonium results.}
\label{tab:charm_beta180}
\end{table}

\end{document}